\documentclass[11pt, a4paper]{article}
\usepackage{styleBuding}
\usepackage{bm}
\usepackage{listings}
\usepackage[usenames,dvipsnames,svgnames,table]{xcolor}
\usepackage{amsmath}
\usepackage{amssymb}
\usepackage{graphicx}
\usepackage{slashed}
\usepackage{soul}
\usepackage{multirow}
\usepackage{subfigure}
\usepackage{siunitx} 
\usepackage[utf8]{inputenc}

\pdfoutput=1

\definecolor{back}{HTML}{F8F8F8}

\usepackage{epstopdf} 

\begin{document}

\title{Singlino-dominated dark matter in $Z_3$-NMSSM}

\author{Haijing Zhou,}
\author{Junjie Cao,}
\author{Jingwei Lian,}
\author{Di Zhang}

\affiliation{Physics Department, Henan Normal University, Xinxiang 453007, China}

\emailAdd{junjiec@alumni.itp.ac.cn}

\abstract{Singlino-dominated dark matter properties are investigated in the $Z_3$ Next-to-Minimal Supersymmetric Standard Model, producing superweak interactions with nucleons involved in dark matter direct detection experiments. Approximate analytical formulas describing the dark matter abundance and cross section in the scattering with nucleons are used to illustrate a dependence on theoretical parameters in neutralino and Higgs sectors. It is shown that the measured abundance requires a sizable singlet--doublet Higgs coupling parameter $\lambda$, while the experimental detection results prefer a small $\lambda$. The parameter space is then surveyed using a nest sampling technique guided by a likelihood function containing various observables in dark matter, Higgs, and B physics, such as the abundance and the scattering cross section. It is demonstrated that dark matter can achieve the correct abundance through $\tilde{\chi}_1^0 \tilde{\chi}_1^0  \to t \bar{t}$ or co-annihilation with higgsinos. The former process provides significantly larger Bayesian evidence than the latter, but this will be examined by the near-future PandaX-4T experiment. If the experiment shows no signs of dark matter, it will become highly disfavored. Furthermore, four cases are summarized to suppress dark matter scattering with nucleons, namely, a small $\lambda$ and three kinds of cancellation between different contributions.}

\maketitle

\section{\label{Introduction}Introduction}

The 2012 discovery of the Higgs boson at the Large Hadron Collider (LHC)~\cite{Aad:2012tfa,Chatrchyan:2012ufa} confirmed the correctness of the Higgs mechanism as the origin of the masses of subatomic particles. However, the hierarchy problem caused by sizeable radiative corrections to the Higgs mass term implies there should be new physics between the electroweak scale and the Planck scale. In addition, the Standard Model (SM) cannot explain the existence of dark matter (DM). However, current astronomical observations have confirmed that the universe is composed of $27\%$ DM~\cite{Ade:2015xua, Aghanim:2018eyx}. Therefore, new DM particle candidates and new physics are required beyond the SM. Establishing the structural nature of DM is one of the most fundamental open questions in cosmology and particle physics.

Among multiple proposed theories, the most widely accepted are supersymmetry (SUSY) models with R-parity conservation, i.e., the minimal supersymmetric standard model (MSSM)~\cite{Haber:1984rc,Gunion:1984yn} and the next-to-minimal supersymmetric standard model (NMSSM)~\cite{Ellwanger:2009dp,Haber:1986gz,Maniatis:2009re}, which provide elegant solutions to the hierarchy problem by introducing contributions from superpartners to the Higgs mass term. In addition, R-parity conservation ensures that the lightest neutralino is a stable neutral particle if it is the lightest supersymmetric particle (LSP), and it may be an excellent DM candidate. MSSM, the most economic realization of SUSY, exhibits several attractive features, but also includes some challenges (e.g., the ``$\mu$-problem''~\cite{Kim:1983dt} and  ``little hierarchy problem''~\cite{BasteroGil:2000bw}) that have become exacerbated in recent years by the first run of the LHC experiments. This was particularly true for the uncomfortably large mass of the discovered Higgs boson $m_h \simeq 125~{\rm GeV}$~\cite{Aad:2015zhl,Hall:2011aa,Ellwanger:2011aa, Gunion:2012zd,King:2012tr, Kang:2012sy,King:2012is,Cao:2012fz,Vasquez:2012hn}. Alternatively, the NMSSM solves the $\mu$-problem by adding a singlet chiral superfield $\hat{S}$ to the MSSM. In this process, the $\mu$ parameter is replaced by a dynamic quantity $\mu_{eff} = \lambda v_s$ when $S$ develops a vacuum expectation value (VEV) $v_s$, the magnitude of which is naturally at the electroweak scale~\cite{Ellwanger:2009dp,Maniatis:2009re}. Furthermore, the SM-like Higgs squared mass can be enhanced by an additional tree-level contribution $\lambda^2 v^2 \sin^2 2 \beta$ and the singlet-doublet Higgs mixing ~\cite{Hall:2011aa,Ellwanger:2011aa,King:2012is,Cao:2012fz}, where $\tan\beta \equiv {v_u}/{v_d}$ with $v_u$ and $v_d$ representing the VEVs of the doublet Higgs fields and $v^2 \equiv v_u^2 + v_d^2$.
As a byproduct, the neutralino sector includes a fermionic partner of $S$ (singlino) in addition to neutral electroweak gauginos (bino and wino) and the neutral fermionic partner of the Higgs doublets (higgsinos). Since the coupling of a singlino field with SM particles may be very weak, this study focuses on the case of a singlino-dominated neutralino as a DM candidate~\cite{Guchait:2020wqn,Das:2012rr,Ellwanger:2014hia,Ellwanger:2016sur,Ellwanger:2018zxt,Aldufeery:2020dkb}. This scenario is feasible when the Yukawa-like couplings satisfy $\lambda \geq 2 \kappa$ and the gauginos are assumed to be heavier than the higgsinos.

In the NMSSM, the Z boson mass is related to the higgsino mass $\mu_{eff}$, and a natural prediction of $m_Z$ favors light higgsinos up to several hundred GeV~\cite{Baer:2012uy}. Generally, the NMSSM with $\mu_{eff} \lesssim 500~\rm{GeV}$ is considered to be a natural NMSSM \cite{King:2012tr,Baer:2012uy,Kang:2012sy,Cao:2014kya,Cao:2016nix}.
However, given the constraints of recent experiments (e.g., the searches for electroweakinos~\cite{Sirunyan:2018lul,Sirunyan:2018nwe,Sirunyan:2018ubx,Sirunyan:2017eie,Sirunyan:2017lae,Sirunyan:2017qaj,Aaboud:2018jiw,Aaboud:2018ngk,Aaboud:2018sua,
Aad:2019vvi,Aad:2019vnb,Aad:2019vvf,ATLAS:2013rla,Sirunyan:2018iwl,TheATLAScollaboration:2013zia,Aad:2014vma,Aad:2014nua,Aad:2015jqa,CMS:2013bda,CMS:2013dea}, the WMAP/Planck experiments~\cite{Ade:2015xua, Aghanim:2018eyx}, and the DM direct detection (DD) and indirect detection (ID) experiments~\cite{Aprile:2018dbl,Aprile:2019dbj,Wang:2020coa,Cui:2017nnn,Ackermann:2015zua}), a large portion of the parameter space in the natural NMSSM has been strongly constrained. As a result, the following isolated and narrow parameter spaces with a singlino-dominated $\tilde \chi^0_1$ are preferred~\cite{Cao:2018rix,Abdallah:2019znp,Abdallah:2020yag,Cao:2019qng}:
\begin{itemize}
\item $\lambda \simeq 2 \kappa$ with $\lambda \lesssim 0.05$, where $\tilde \chi^0_1$ mainly co-annihilates with higgsinos to achieve the measured abundance~\cite{Cao:2018rix}.
\item $\kappa \sim 0.01$, $\lambda \lesssim 0.2$, and at least one light singlet-dominated Higgs boson~\cite{Abdallah:2019znp}. Here, $\tilde \chi^0_1$ annihilates in certain funnel regions and the higgsinos decay in a complex manner to satisfy the LHC constraints.
\end{itemize}
These conclusions are applicable for $|\mu_{eff}| \lesssim 500~{\rm GeV}$, or equivalently, the fine-tuning criterion ${ {\Delta}m_Z} \lesssim 50$, where $\Delta m_Z$ defined in~\cite{Ellwanger:2011mu} parameterizes the sensitivity of $m_Z$ to the SUSY parameters at the weak scale. Given this situation and the fact that the fine-tuning criteria lacks a confirmed scientific basis and may reflect personal prejudice, we update the study by Cao et al. \cite{Cao:2018rix} to improve these conclusions. Specifically, we do not require ${\Delta}m_Z$ to be less than 50 but impose the condition that $\mu_{eff} \leq 1~{\rm TeV}$. We adopt an advanced MultiNest algorithm~\cite{Feroz:2008xx,Feroz:2013hea} to perform a sophisticated scan over the parameter space of the $Z_3$-invariant NMSSM with a singlino-dominated DM. This algorithm is much more efficient than the other algorithms (e.g., the Markov Chain method~\cite{Markov-Chain} adopted in~\cite{Cao:2018rix}) in providing comprehensive information of the space to reveal the underlying physics, although it usually involves a tremendous amount of calculation. To the best of our knowledge, few researchers have used it to study the NMSSM phenomenology~\cite{Gunion:2011hs,Kowalska:2012gs,Kim:2013uxa,Roszkowski:2014lga,Athron:2017fxj,AbdusSalam:2017uzr}. We also present a description of DM annihilation and the mechanisms used to suppress the DM nucleon scattering cross section through analytical formulas and numerical analysis. Some of the formulas are new, and some are consistent with excellent related works ~\cite{Cheung:2014lqa,Cao:2015loa,Badziak:2015exr,Baum:2017enm,Badziak:2017uto}. Evidently, such an analysis is helpful to understand the DM physics.

The remainder of this paper is organized as follows. In Section~\ref{sec:model}, we briefly introduce the basic properties of the $Z_3$-invariant NMSSM, including the Higgs and neutralino sections. We then demonstrate DM annihilation and scattering cross sections for singlino-dominated $\tilde{\chi}^0_{1}$ with nucleons using analytical formulas. In Section~\ref{sec:scan}, we provide a brief description of our scanning strategy. In Section~\ref{result13}, we investigate predictions for surviving samples and the properties of singlino-dominated DM scenarios to understand their distinctive features. Finally, Section~\ref{sec:conclusion} includes a discussion of the results and corresponding conclusions.

\section{\label{sec:model} Next-to-minimal supersymmetric standard model model}

\subsection{Fundamental NMSSM properties}

As the simplest extension of the MSSM, the NMSSM includes one additional gauge singlet Higgs field $\hat{S}$. The associated superpotential can be expressed as follows~\cite{Maniatis:2009re,Ellwanger:2009dp}:
\begin{align}
\label{eq:superpotential}
  W_\mathrm{NMSSM}=W_\mathrm{MSSM} + \lambda \hat{S} \hat{H_u} \hat{H_d} + \frac{1}{3} \kappa \hat{S}^3,
\end{align}
where $W_\mathrm{MSSM}$ is the MSSM superpotential without the $\mu$ term, $\lambda$ and $\kappa$ are dimensionless parameters, and $\hat{H}_u$ and $\hat{H}_d$ are the common Higgs superfields. It is the most general R-parity-conservation superpotential satisfying a $Z_3$ discrete symmetry given the considered field content.

Assuming CP-conservation, the Higgs sector of the $Z_3$-NMSSM is determined by six parameters at the tree-level~\cite{Ellwanger:2009dp,Cheung:2010ba}:
\begin{align}
  \lambda,~ \kappa,~ A_\lambda,~ A_\kappa,~ \mu_{eff},~ \tan\beta,
  \label{eq:six}
\end{align}
where $A_\lambda$ and $A_\kappa$ are the soft trilinear coefficients defined in Eq. (2.5) of Ref.~\cite{Ellwanger:2009dp}.
In the  base-vectors, $H_{\rm SM} \equiv \sin{\beta} Re[H_{u}^0] + \cos{\beta} Re[H_{d}^0], H_{\rm NSM} \equiv \cos{\beta} Re[H_{u}^0] - \sin{\beta} Re[H_{d}^0]$, and $H_{\rm S} \equiv Re[S]$ for CP-even fields and $A_{\rm NSM} \equiv \cos{\beta} Im[H_{u}^0] + \sin{\beta} Im[H_{d}^0]$ and $A_{\rm S}\equiv Im[S])$ for CP-odd fields\footnote{$H_u^0$, and $H_d^0$ denote the neutral component fields of the doublet scalar fields $H_u$ and $H_d$, respectively.}, the three CP-even mass eigenstates $h_i = \{ h, H, h_s \}$ and two CP-odd Higgs mass eigenstates $a_i = \{A_H, a_s\}$ are given as follows:
\begin{eqnarray}\label{eq:hi}
h_i &=&  V_{h_i}^{\rm SM} H_{\rm SM}+V_{h_i}^{\rm NSM} H_{\rm NSM}+V_{h_i}^{\rm S} H_{\rm S}, \nonumber \\
a_i &=&  V_{a_i}^{' \rm NSM} A_{\rm NSM} + V_{a_i}^{'\rm S} A_S,
\end{eqnarray}
where $V$ and $V'$ represent the unitary matrices to diagonalize the corresponding Higgs squared mass matrix. In this work, we denote the physical Higgs state with the largest $H_{\rm SM}$ component by the symbol $h$, which is called the SM-like Higgs boson hereafter, and we denote the physical Higgs state with the largest non-SM doublet (singlet) component $H_{\rm NSM}$ ($H_{\rm S}$) by  $H$ ($h_{s}$). We also denote the CP-even Higgs bosons by $h_1$, $h_2$, and $h_3$, with $m_{h_1} < m_{h_2} < m_{h_3}$. The latter notation is primarily for convenience. To date, the LHC experiments have measured the couplings of the discovered Higgs boson with about $10\%$ uncertainty, and they revealed that the boson has roughly the same couplings as the SM Higgs boson~\cite{Aad:2019mbh,Sirunyan:2018koj}. These facts imply that $\sqrt{\left (V_{h}^{\rm NSM} \right )^2 + \left ( V_{h}^{\rm S} \right )^2} \lesssim 0.1$ and $|V_{h}^{\rm SM}| \sim 1$.

In the $Z_3$-NMSSM, mixtures of bino ($\tilde{B}^0$), wino ($\tilde{W}^0$), higgsino ($\tilde{H}_{d,u}^0$), and
singlino ($\tilde{S}^0$) fields form neutralinos. Assuming a basis of $\psi^0 = (-i \tilde{B}, - i \tilde{W}^0, \tilde{H}_{d}^0, \tilde{H}_{u}^0,\tilde{S})$ produces the following neutralino mass matrix~\cite{Ellwanger:2009dp}:
\begin{align}
{\cal M} = \left(
\begin{array}{ccccc}
M_1 & 0 & -\frac{g_1 v_d}{\sqrt{2}} & \frac{g_1 v_u}{\sqrt{2}} & 0 \\
  & M_2 & \frac{g_2 v_d}{\sqrt{2}} & - \frac{g_2 v_u}{\sqrt{2}} &0 \\
& & 0 & -\mu_{eff} & -\lambda v_u \\
& & & 0 & -\lambda v_d\\
& & & & \frac{2 \kappa}{\lambda} \mu_{eff}
\end{array}
\right), \label{eq:massmatrix}
\end{align}
where $M_1$, $M_2$, and $\mu_{eff}$ denote the soft breaking masses of the bino, wino, and higgsinos, respectively. Diagonalizing the mass matrix with a unitary matrix $N$ yields five mass eigenstates (ordered by mass):
\begin{align}\label{eq:chi01all}
\tilde{\chi}_i^0 =  N_{i1} \tilde B^0 + N_{i2} \tilde W_3^0 + N_{i3}\tilde{H}_d^0 + N_{i4}\tilde{H}_u^0 + N_{i5}\tilde{S}.
\end{align}
The lightest neutralino, $\tilde{\chi}_1^0$, acting as the DM candidate is the focus of this work.

In the limit that $|M_1|$ and $|M_2|$ are much larger than $|\mu_{eff}|$ and $v$, $\tilde{\chi}_1^0$ is approximated by
\begin{align}\label{eq:chi01SH}
\tilde{\chi}_1^0 \simeq  N_{13}\tilde{H}_d^0 + N_{14}\tilde{H}_u^0 + N_{15}\tilde{S}.
\end{align}
If $|\kappa/\lambda| < 1$, the dominant composition of $\tilde{\chi}_1^0$ is a singlino. In this case,
$\kappa$ is related to $m_{\tilde{\chi}_1^0}$ as follows~\cite{Cheung:2014lqa,Cao:2015loa}:
\begin{eqnarray}
\kappa & = & \frac{\lambda}{2 \mu_{eff}} \left[ m_{\tilde{\chi}_1^0}
  - \frac{\lambda^2 v^2 \left( m_{\tilde{\chi}_1^0} - \mu_{eff} \sin 2\beta
    \right)}{m_{\tilde{\chi}_1^0}^2 - \mu_{eff}^2} \right], \label{eq:kappa}
\end{eqnarray}
and the elements of the matrix $N$ exhibit the following relationships~\cite{Cheung:2014lqa,Badziak:2015exr,Badziak:2017uto,Baum:2017enm,Cao:2015loa}:
%
\begin{align}
\label{eq:N13N15}
\frac{N_{13}}{N_{15}}
=
\frac{\lambda v}{\mu_{eff}}
\,
\frac{(m_{\tilde{\chi}_1^0}/\mu_{eff})\sin\beta-\cos\beta}
{1-\left(m_{\tilde{\chi}_1^0}/\mu_{eff} \right)^2}
\,,\\[4pt]
\label{eq:N14N15}
\frac{N_{14}}{N_{15}}
=
\frac{\lambda v}{\mu_{eff}}
\,
\frac{(m_{\tilde{\chi}_1^0}/\mu_{eff})\cos\beta-\sin\beta}
{1-\left(m_{\tilde{\chi}_1^0}/\mu_{eff} \right)^2}.
\end{align}
Thus,
\begin{eqnarray}
\label{eq:N15}
N_{15}^2 && \simeq \left(1+ \frac{N^2_{13}}{N^2_{15}}+\frac{N^2_{14}}{N^2_{15}}\right)^{-1}  \\
&&\simeq \frac{\left[1-(m_{\tilde{\chi}_1^0}/\mu_{eff} )^2\right]^2}{\left[(m_{\tilde{\chi}_1^0}/\mu_{eff} )^2
-2(m_{\tilde{\chi}_1^0}/\mu_{eff} )\sin2\beta+1 \right]\left({\lambda v}/{\mu_{eff}}\right)^2
+\left[1-(m_{\tilde{\chi}_1^0}/\mu_{eff})^2\right]^2}.  \nonumber
\end{eqnarray}
The higgsino and singlino fractions in $\tilde{\chi}_1^0$ can be defined as
$Z_h=N_{13}^2+N_{14}^2$ and $Z_s=N_{15}^2$, respectively. The ratio of $Z_h$ to $Z_s$ can then be expressed as follows:
\begin{align}
\label{eq:Higgsino/singlino}
\frac{Z_h}{Z_s}
=
\left(\frac{\lambda v}{\mu_{eff}}\right)^{\!\!2}
\frac{\left(m_{\tilde{\chi}_1^0}/\mu_{eff}\right)^2-2{(m_{\tilde{\chi}_1^0}}/{\mu_{eff}})\sin2\beta+1}
{\left[1-\left({m_{\tilde{\chi}_1^0}}/{\mu_{eff}}\right)^2\right]^2}
\,.
\end{align}
This expression implies that a small $\lambda$ can suppress the higgsino fraction in $\tilde{\chi}_1^0$.

The couplings of DM to scalar Higgs states, the Z boson, and the Goldstone boson $G^0$ are included in the calculation of DM annihilation.
They take the following form~\cite{Ellwanger:2009dp}:
\begin{eqnarray}
{\cal{L}}_{\rm NMSSM} \ni C_{\tilde{\chi}_1^0 \tilde{\chi}_1^0 Z} Z_\mu \overline{\tilde{\chi}_1^0} \gamma^\mu \gamma_5 \tilde{\chi}_1^0 + i C_{\tilde{\chi}_1^0 \tilde{\chi}_1^0 G^0} G^0 \overline{\tilde{\chi}_1^0} \gamma_5 \tilde{\chi}_1^0 + C_{\tilde{\chi}_1^0 \tilde{\chi}_1^0 h_i} h_i \overline{\tilde{\chi}_1^0} \tilde{\chi}_1^0 + i C_{\tilde{\chi}_1^0 \tilde{\chi}_1^0 a_i} a_i \overline{\tilde{\chi}_1^0} \gamma_5 \tilde{\chi}_1^0, \nonumber
\end{eqnarray}
where the coefficients are given by
\begin{align}
C_{\tilde{\chi}_1^0 \tilde{\chi}_1^0 Z}\simeq & \frac{m_Z}{\sqrt{2} v} \big( \frac{\lambda v}{\mu_{eff}} \big)^2
\,
\frac{Z_s\cos2\beta}
{1-(m_{\tilde{\chi}_1^0}/\mu_{eff})^2},
\label{eq1:zchi10chi10_S} \\
C_{\tilde {\chi}^0_1 \tilde {\chi}^0_1 G^0 }
\simeq&  \frac{\sqrt{2} \mu_{eff}}{v} \left( \frac{\lambda v}{\mu_{eff}}\right)^2 \frac{Z_s(m_{\tilde{\chi}_1^0}/\mu_{eff})\cos2\beta}{1-(m_{\tilde{\chi}_1^0}/\mu_{eff})^2},
\label{eq1:G0chi10chi10_S}
\\
C_{\tilde {\chi}^0_1 \tilde {\chi}^0_1 h_{i}}  \simeq &
V_{h_i}^{\rm SM} C_{\tilde {\chi}^0_1 \tilde {\chi}^0_1 H_{\rm SM} }+V_{h_i}^{\rm NSM} C_{\tilde {\chi}^0_1 \tilde {\chi}^0_1 H_{\rm NSM} }+V_{h_i}^{\rm S} C_{\tilde {\chi}^0_1 \tilde {\chi}^0_1 H_{\rm S} } \nonumber \\
 \simeq &
\frac{\sqrt{2}\mu_{eff}}{v}\,\big( \frac{\lambda v}{\mu_{eff}} \big)^2\, \frac { Z_s V_{h_{i}}^{\rm SM}(m_{\tilde{\chi}_1^0}/\mu_{eff} -\sin 2 \beta)}{1-(m_{\tilde{\chi}_1^0}/\mu_{eff})^2}  - \frac{\sqrt{2}\mu_{eff}}{v}\,\big( \frac{\lambda v}{\mu_{eff}} \big)^2\, \frac { Z_s V_{h_{i}}^{\rm NSM}\cos 2 \beta}{1-(m_{\tilde{\chi}_1^0}/\mu_{eff})^2}   \nonumber \\
&+ \lambda \big( \frac{\lambda v}{\mu_{eff}} \big)^2 \frac{Z_s V_{h_{i}}^{\rm S} \sin2\beta}{\sqrt{2}\big[ 1-(m_{\tilde{\chi}_1^0}/\mu_{eff})^2 \big]}-\sqrt{2}\kappa Z_s V_{h_{i}}^{\rm S} \left[1+ \big( \frac{\lambda v}{\mu_{eff}} \big)^2\frac{2}{1-(m_{\tilde{\chi}_1^0}/\mu_{eff})^2}  \right],
\label{eq1:hichi01chi01_S}
\\
C_{\tilde {\chi}^0_1 \tilde {\chi}^0_1 a_{i}} \simeq &
V_{a_i}^{' \rm NSM}C_{\tilde {\chi}^0_1 \tilde {\chi}^0_1 A_{\rm NSM} }+ V_{a_i}^{' \rm S}C_{\tilde {\chi}^0_1 \tilde {\chi}^0_1 A_{\rm S} } \nonumber \\
 \simeq &
-\frac{\sqrt{2}\mu_{eff}}{v}\,\big( \frac{\lambda v}{\mu_{eff}} \big)^2\, \frac { Z_s V_{a_{i}}^{' \rm NSM}(m_{\tilde{\chi}_1^0}/\mu_{eff} \sin 2 \beta-1)}{1-(m_{\tilde{\chi}_1^0}/\mu_{eff})^2}  \nonumber \\
&+ \lambda \big( \frac{\lambda v}{\mu_{eff}} \big)^2 \frac{Z_s V_{a_{i}}^{'\rm S} \sin2\beta}{\sqrt{2}\big[ 1-(m_{\tilde{\chi}_1^0}/\mu_{eff})^2 \big]}
-\sqrt{2}\kappa Z_s V_{a_{i}}^{'\rm S} \left[1+ \big( \frac{\lambda v}{\mu_{eff}} \big)^2\frac{2}{1-(m_{\tilde{\chi}_1^0}/\mu_{eff})^2}  \right],
\label{eq1:aichi01chi01_S}
\end{align}
where the approximation in Eq. (\ref{eq:chi01SH}) was applied. In addition, the coupling of a CP-even Higgs $h_i$ to two quasi-pure singlet CP-odd Higgs $a_s$,
${\cal{L}}_{\rm NMSSM} \ni C_{h_i a_s a_s} h_i a_s a_s$, is relevant to our study. Its coefficient $C_{h_i a_s a_s}$ is given
as follows~\cite{Ellwanger:2004xm,Belanger:2005kh}:
\begin{align}
C_{h_i a_s a_s}\simeq& \sqrt{2}\lambda v\,V_{h_i}^{\rm SM} \,(\lambda+\kappa \sin2\beta)+ \sqrt{2}\lambda\kappa v\,
 V_{h_i}^{\rm NSM}\cos2\beta
+\sqrt{2} V_{h_i}^{\rm S}( 2 \kappa^2 v_s -\kappa A_\kappa).
\label{eq:hiasas}
\end{align}
In most cases, $C_{h_s a_s a_s} \gg C_{h a_s a_s}$, since $v_s \gg v$.

\subsection{Dark matter relic density}

In the NMSSM, the abundance of the singlino-dominated DM candidate ($\tilde \chi^0_1$) tends to be unacceptably large, due to small coupling effects with SM particles. However, such a candidate can still achieve the measured abundance~\cite{Ade:2015xua, Aghanim:2018eyx} by a specific mechanism, e.g., via $s$-channel exchanges of gauge and Higgs bosons and $t$-channel exchanges of electroweakinos and sfermions. It can also be achieved through co-annihilation with heavier states, such as sleptons, the next-to-lightest neutralino ($\tilde{\chi}_2^0$), or the lightest chargino ($\tilde{\chi}_1^\pm$).

The thermal abundance of the DM at the freeze-out temperature $T_F = m_{\tilde{\chi}_1^0}/x_F$ is given as follows~\cite{Baum:2017enm}:
\begin{equation}\label{density}
\Omega h^2 = 0.12\left(\frac{80}{g_*}\right)^{1/2}\left(\frac{x_F}{25}\right) \left( \frac{2.3\times 10^{-26}\ \mathrm{cm^3/s}}{\langle \sigma v\rangle_{x_F}}\right)\;,
\end{equation}
with a thermally-averaged annihilation cross section ${\langle \sigma v\rangle_{x_F}} \equiv a+ \frac{3b}{x_F}$. Dominant contributions to $\langle \sigma v \rangle_{x_F}$ in acquiring the measured abundance are discussed in Section 2.2 of Baum et al.~\cite{Baum:2017enm}. The following conclusions were presented:

1) $\tilde{\chi}_1^0 \tilde{\chi}_1^0 \to t\bar{t}$ is usually the most crucial channel for the abundance when $m_{\tilde{\chi}_1^0} > m_t$. It proceeds through the s-channel exchanges of Higgs and Z bosons. Since the top quark is massive, the contribution from the Z boson's longitudinal polarization to $\langle \sigma v \rangle_{x_F}$ is important in this process when the Higgs mediators are far off-shell. $\langle \sigma v \rangle_{x_F}$ can then be approximately expressed as follows:
\begin{equation} \label{eq:gxxGforOmega}
\langle \sigma v \rangle_{x_F}^{t\bar{t}} \sim 2 \times 10^{-26}\,\frac{{\rm cm}^3}{{\rm s}} \left( \frac{\left|C_{\tilde{\chi}_1^0 \tilde{\chi}_1^0  G^0}\right|}{0.1} \right)^2 \left( \frac{m_{\tilde{\chi}_1^0} }{300\,{\rm GeV}}\right)^{-2}.
\end{equation}
The measured abundance $\Omega h^2 \sim 0.12$~\cite{Ade:2015xua,Aghanim:2018eyx} is achieved through the coupling of DM pairs to the Goldstone boson $|C_{\tilde{\chi}_1^0 \tilde{\chi}_1^0  G^0}| \sim 0.1 $, which requires $\lambda \gtrsim 0.4$ according to Eq. (\ref{eq1:G0chi10chi10_S}).

2) $\tilde{\chi}_1^0 \tilde{\chi}_1^0 \to \Phi_i \Phi_j$ is another crucial annihilation process for the abundance, where $\Phi_i$ denotes a scalar or pseudo-scalar Higgs mass eigenstate. Such processes occur via $s$-channel Higgs or $Z$ boson exchange and $t$-channel neutralino exchange. Since $\tilde{\chi}_1^0 \tilde{\chi}_1^0 \to h_i h_j$ and $\tilde{\chi}_1^0 \tilde{\chi}_1^0 \to a_i a_j$ are $p$-wave suppressed~\cite{Baum:2017enm}, and because $C_{\tilde{\chi}_1^0 \tilde{\chi}_1^0 h_i}$ with $h_i = h, H$ in Eq.~(\ref{eq1:hichi01chi01_S}) and
$C_{\tilde{\chi}_1^0 \tilde{\chi}_1^0 A_H}$ in Eq.~(\ref{eq1:aichi01chi01_S}) are usually smaller than 0.1, here we consider only the contribution from $\tilde{\chi}_1^0 \tilde{\chi}_1^0 \to h_s a_s$ to $\langle \sigma v \rangle_{x_F}$.  $\langle \sigma v \rangle_{x_F}^{h_s a_s}$ is then given as follows~\cite{Baum:2017enm,Griest:1990kh}:
\begin{eqnarray} \label{eq:sigvPhiPhi}
\left\langle \sigma v \right\rangle_{x_F}^{h_s a_s} \simeq && \frac{1}{64 \pi m_{\tilde{\chi}_1^0}^2} \left\{ \left[1-\frac{\left(m_{h_s} + m_{a_s}\right)^2}{4 m_{\tilde{\chi}_1^0}^2}\right] \left[1-\frac{\left(m_{h_s} - m_{a_s}\right)^2}{4 m_{\tilde{\chi}_1^0}^2}\right] \right\}^{1/2} | {\cal{A}}_s + {\cal{A}}_t |^2, \quad \label{hsaa-approximation}
\end{eqnarray}
where the $s$- and $t$-channel contributions are approximated as
\begin{eqnarray}
{\cal{A}}_s &\simeq & \frac{-2 m_{\tilde{\chi}_1^0} C_{\tilde {\chi}^0_1 \tilde {\chi}^0_1 a_s }  C_{h_s a_s a_s}}{m_{a_s}^2 - 4  m_{\tilde{\chi}_1^0}^2}, \nonumber \\
{\cal{A}}_t &\simeq & - 2 C_{\tilde{\chi}_1^0\tilde{\chi}_1^0 h_s} \, C_{\tilde{\chi}_1^0\tilde{\chi}_1^0 a_s} \left[ 1 + \frac{ 2 m_{a_s}^2}{ 4 m_{\tilde{\chi}_1^0}^2 - \left(m_{h_s}^2 + m_{a_s}^2\right) } \right],
\end{eqnarray}
if there are no resonant contributions\footnote{Note that the $\tilde{\chi}_i^0$-mediated ($i \neq 1$) contribution to $\langle \sigma v \rangle_{x_F}^{h_s a_s}$ is less important than the $\tilde{\chi}_1^0$-mediated contribution for two reasons. One is that, if $|\kappa|$ is comparable to $\lambda$, $|C_{\tilde{\chi}_1^0 \tilde{\chi}_i^0 h_s}|$ and $|C_{\tilde{\chi}_1^0 \tilde{\chi}_i^0 a_s}|$ are significantly smaller than $|C_{\tilde{\chi}_1^0 \tilde{\chi}_1^0 h_s}|$ and $|C_{\tilde{\chi}_1^0 \tilde{\chi}_1^0 a_s}|$, respectively. The other is that, since $m_{\tilde{\chi}_i^0} > m_{\tilde{\chi}_1^0}$, the former contribution is relatively suppressed by the propagator.}. According to Eqs. (\ref{eq1:hichi01chi01_S}) and (\ref{eq1:aichi01chi01_S}), if $h_s$, $a_s$, and  $\tilde{\chi}_1^0$ are pure singlet states, then $|C_{ \tilde{\chi}_1^0 \tilde{\chi}_1^0 h_i}|  =  |C_{\tilde{\chi}_1^0 \tilde{\chi}_1^0 a_i }| \sim \sqrt{2}|\kappa|$. The measured abundance then requires $\kappa \sim 0.15 \left( \frac{m_{\tilde{\chi}_1^0}}{300\,{\rm GeV}}\right)^{1/2} $ in the case of $|{\cal{A}}_t| \gg |{\cal{A}}_s|$ ~\cite{Baum:2017enm}.
Hence, it is evident that once the involved particle masses are fixed, the density is primarily determined by the parameter $\kappa$. However, because $\lambda > 2 |\kappa|$ to ensure a singlino-dominated $\tilde{\chi}_1^0$, the measured abundance can set a lower bound on $\lambda$.

To date, the sensitivities of the XENON-1T experiments have reached the precision of $10^{-47}~{\rm cm^2}$ for the SI cross section~\cite{Aprile:2018dbl} and $10^{-42}~{\rm cm^2}$ for the SD cross section~\cite{Aprile:2019dbj}. They have strongly restricted the $\lambda \gtrsim 0.3$ case (see the discussion about DM-nucleon scattering), which is preferred by the annihilations $\tilde{\chi}_1^0 \tilde{\chi}_1^0 \to t \bar{t}, h_s a_s$ to account for the measured abundance. In addition, the LHC searches for new particles~\cite{Sirunyan:2018lul,Sirunyan:2018nwe,Sirunyan:2018ubx,Sirunyan:2017eie,Sirunyan:2017lae,Sirunyan:2017qaj,Aaboud:2018jiw,Aaboud:2018ngk,Aaboud:2018sua,
Aad:2019vvi,Aad:2019vnb,Aad:2019vvf,ATLAS:2013rla,Sirunyan:2018iwl,TheATLAScollaboration:2013zia,Aad:2014vma,Aad:2014nua,Aad:2015jqa,CMS:2013bda,CMS:2013dea}, and its precise measurement of the discovered scalar's properties~\cite{Aad:2019mbh,Sirunyan:2018koj} have limited the theoretical prediction on light sparticles and Higgs bosons. This has a significant impact on the DM annihilation channels since they are usually accompanied with light particles to account for the abundance (see the discussion in~\cite{Cao:2018rix}). Thus, the scenario preferred by the scan of Ref.~\cite{Cao:2018rix} involves a singlino-dominated DM with $\lambda \lesssim 0.05$. In this case, an effective mechanism to obtain the measured abundance includes co-annihilation with higgsinos. The corresponding reaction is $\tilde{\chi}_i \tilde{\chi}_j \rightarrow X X^\prime $, in which $ XX^\prime$ denotes SM particles and  $\tilde{\chi}_i \tilde{\chi}_j$ may be an LSP-NLSP or NLSP-NLSP annihilation state (e.g., $\tilde{\chi}_1^0 \tilde{\chi}^0_2$,  $\tilde{\chi}_1^0 \tilde{\chi}^+_1$, or $\tilde{\chi}_2^0 \tilde{\chi}^+_1$). This mechanism is distinct in that the effective annihilation rate at a temperature $T$ is very sensitive to the $\tilde{\chi}_1^0$-higgsino mass splitting~\cite{Griest:1990kh,Baker:2015qna}, and even for a small $\lambda$ and $\kappa$, it can still explain the measured abundance\footnote{Note that the co-annihilation mechanism applies under the premise that $\tilde{\chi}_1^0$ and the higgsinos remained in thermal equilibrium in the early universe~\cite{Griest:1990kh,Baker:2015qna}. In the $Z_3$-NMSSM, many processes, such as $\tilde{\chi}_1^0 \tilde{\chi}_1^0 \leftrightarrow \tilde{\chi}_i \tilde{\chi}_j$, $\tilde{\chi}_1^0 X \leftrightarrow \tilde{\chi}_i X^\prime$, and $\tilde{\chi}_i \leftrightarrow \tilde{\chi}_1^0 X X^\prime$, could keep $\tilde{\chi}_1^0$ in chemical equilibrium with $\tilde{\chi}_i$, and the conversion rates of some of them might be enhanced if the mediator were around its mass-shell. We add that maintaining the thermal equilibrium does not necessarily require the involved couplings to be moderately large. For example, the equilibrium condition was discussed in Eqs. (4.3) and (4.4) of Ref.~\cite{Baker:2015qna} in the framework of the DM model ST11. It was found that the involved coupling may be as low as $10^{-4}$ to maintain the equilibrium.}.

\subsection{DM-nucleon cross sections}

Serving for a WIMP, $\tilde {\chi}^0_1$ might be detected by measuring the recoil of a nucleus after an elastic scattering of $\tilde {\chi}^0_1$ on a nucleus taking place. In the non-relativistic limit, only two different kinds of interactions between a neutralino and a nucleon  need to be considered~\cite{Jungman:1995df}: the spin-dependent interaction (SD) where the WIMP couples to the spin of the nucleus, and the spin-independent interaction (SI) where the WIMP couples to the mass of the nucleus.

When $m_{\tilde{q}} \gtrsim 2~{\rm TeV}$, only the $t$-channel $Z$ exchange diagram contributes significantly to the spin-dependent (SD) scattering cross section at the tree level, which is approximated by~\cite{Pierce:2013rda,LDM-27}
\begin{align}
\label{eq:sigSD}
\sigma_{\tilde{\chi}_1^0-N}^{\rm SD} \simeq  C_N \times \left ( \frac{C_{\tilde{\chi}_1^0 \tilde{\chi}_1^0 Z}}{0.01} \right )^2,
\end{align}
with $N=p(n)$ denoting protons (neutrons) and $C_p \simeq 2.9 \times 10^{-41}~{\rm cm^2} $ ($C_n \simeq 2.3 \times 10^{-41}~{\rm cm^2} $)~\cite{Badziak:2015exr,Badziak:2017uto}.
From the expression for $C_{\tilde{\chi}_1^0 \tilde{\chi}_1^0 Z}$ in Eq. (\ref{eq1:zchi10chi10_S}), it is evident that $\sigma_{\tilde{\chi}_1^0-N}^{\rm SD}$ is proportional to $(\lambda v /\mu_{eff})^4$. Furthermore, in the co-annihilation case, the degeneracy of $m_{\tilde{\chi}_1^0}$ and $\mu_{eff}$ leads to a minuscule denominator in Eq. (\ref{eq1:zchi10chi10_S}), which requires a small value of $\lambda v /\mu_{eff}$ to satisfy the DM-DD experimental constraints.

In contrast, the spin-independent (SI) scattering cross section in the heavy squark limit is dominated by a $t$-channel exchange of CP-even Higgs bosons $h_i$~\cite{Drees1993,Drees1992,Jungman1995,Belanger2008} and can be expressed as~\cite{cao:2021Gnmssm}
\begin{equation} \label{eq:SIDD_p}
	\sigma_{\tilde {\chi}^0_1-{N}}^{\rm SI} = \frac{ m_N^2}{2\pi v^2} \left( \frac{m_N m_{\tilde{\chi}_1^0}}{m_N + m_{\tilde{\chi}_1^0}} \right)^2 {\left( \frac{1}{125~\rm GeV} \right)^4}\left\{ \sum_{h_i} \left[ F^N_u \left({a_u}\right)_{h_i} + F^N_d \left({a_d}\right)_{h_i} \right] \right\}^2,
\end{equation}
where $m_N$ is the nucleon mass, $F^{(N)}_d=f^{(N)}_d+f^{(N)}_s+\frac{2}{27}f^{(N)}_G$ and $F^{(N)}_u=f^{(N)}_u+\frac{4}{27}f^{(N)}_G$ with $f^{(N)}_q =m_N^{-1}\left<N|m_qq\bar{q}|N\right> $ ($q=u,d,s$) represent the normalized light quark contribution to the nucleon mass, and $f^{(N)}_G=1-\sum_{q=u,d,s}f^{(N)}_q$ influences other heavy quark mass fractions in nucleons~\cite{Drees1993,Drees1992}. In this study, the default settings for $f_q^{N}$ were used in the micrOMEGAs package~\cite{Belanger2008}, and they predict $F_u^{p} \simeq F_u^n \simeq 0.15$ and $F_d^{p} \simeq F_d^n \simeq 0.13$. Hence, SI cross sections for DM-proton scattering and DM-neutron scattering are approximately equal (i.e., $\sigma^{SI}_{{\tilde {\chi}^0_1}-p} \simeq \sigma^{SI}_{{\tilde {\chi}^0_1}-n}$)~\cite{DM-detecion-SI-SD}. The quantities
$\left( a_u\right)_{h_i}$ and $\left( a_d\right)_{h_i}$ are defined by
\begin{eqnarray}
	\left( {a_u} \right)_{h_i} &=&  \left( \frac{125~\rm GeV}{m_{h_i} }\right)^2 \left(V_{h_i}^{\rm SM}+\frac{1}{\tan\beta} V_{h_i}^{\rm NSM} \right) C_{ \tilde {\chi}^0_1 \tilde {\chi}^0_1 h_i} ~, \\
	\left( {a_d}\right)_{h_i} &=& \left(  \frac{125~\rm GeV}{ m_{h_i}}\right)^2 \left( V_{h_i}^{\rm SM}- \tan\beta V_{h_i}^{\rm NSM}\right)  C_{ \tilde {\chi}^0_1 \tilde {\chi}^0_1 h_i} ~.
\end{eqnarray}

Currently, non-SM doublet Higgs bosons $H$ are preferred to be heavier than several hundreds of GeV in LHC experiments. In this case, the contribution from $H$ to the SI cross section will be suppressed by $(a_q)_{H}^2 \propto 1/m_{H}^4$, and it is much smaller than that from $h$ for a not exceedingly large $\tan \beta$. As such, the primary contribution to SI scattering comes from the t-channel exchange of SM-like Higgs bosons $h$ and the singlet Higgs boson $h_s$. The latter contribution may be crucial when $h_s$ is much lighter than $h$. Since the non-SM doublet components of $h$ and $h_s$ are approximately zero, $\left( {a_u} \right)_{h} + \left( {a_u} \right)_{h_{s}} \simeq \left( {a_d}\right)_{h} + \left( {a_d}\right)_{h_{s}}
\equiv \mathcal{A }$, which can be expressed as
\begin{align}\label{eq0:au-ad_hsm-hs}
\mathcal{A} \simeq \left( \frac{125 \rm GeV}{m_{h}}\right)^2 V_{h}^{\rm SM} C_{ \tilde {\chi}^0_1 \tilde {\chi}^0_1 h} +  \left( \frac{125 \rm GeV}{m_{h_{s}}} \right)^2  V_{h_{s}}^{\rm SM} C_{ \tilde {\chi}^0_1 \tilde {\chi}^0_1 h_{s}}~,
\end{align}
where $C_{ \tilde {\chi}^0_1 \tilde {\chi}^0_1 h}$ and $C_{ \tilde {\chi}^0_1 \tilde {\chi}^0_1 h_{s}}$ are given by Eq. (\ref{eq1:hichi01chi01_S}).
Thus, the SI scattering cross section in Eq. (\ref{eq:SIDD_p}) can be rewritten as
\begin{align}
\sigma_{\tilde {\chi}^0_1-{N}}^{\rm SI}  &\simeq \frac{m_N^2}{2 v^2\pi} \left( \frac{m_N m_{\tilde{\chi}_1^0}}{m_N + m_{\tilde{\chi}_1^0}} \right)^2 {\left( \frac{1}{125 \rm GeV} \right)^4}  \left( F^N_u  + F^N_d  \right)^2 \mathcal{A}^2\sim  5 \times 10^{-45} {\rm cm^2}\times \left(\frac{\mathcal{A}}{0.1}\right)^2.
\label{eq:SIDD_p2}
\end{align}
In these expressions, if only the contribution from a pure SM Higgs state is considered, $\mathcal{A}$ can be simply expressed as
\begin{align}\label{eq0:au-ad_hsm}
 \mathcal{A} \sim  \left( \frac{125 \rm GeV}{m_{h}}\right)^2 \frac{\sqrt{2}\lambda^2 v}{\mu_{eff}} \, \frac { Z_s (m_{\tilde{\chi}_1^0}/\mu_{eff} -\sin 2 \beta)}{1-(m_{\tilde{\chi}_1^0}/\mu_{eff})^2}.
\end{align}
It is immediately evident that $\sigma_{\tilde{\chi}_1^0-N}^{\rm SI}$ will vanish for $m_{\tilde{\chi}_1^0}/\mu_{eff}=\sin2\beta$, which corresponds to a blind spot (BS) condition in~\cite{Badziak:2015exr, Badziak:2016qwg, Baum:2017enm}.

The above analytical formulas for the SD and SI scattering cross sections suggest that $\sigma^{SD} \varpropto \big( \frac{\lambda v}{\mu_{eff}} \big)^2\frac{1}{1-(m_{\tilde{\chi}_1^0}/\mu_{eff})^2}$ and (by contrast) $\sigma^{SI}$ depends on $\lambda$, $\mu_{eff}$, and $m_{\tilde{\chi}^0_1}$ in a complex way. In general, a larger $\lambda$, a smaller $\mu_{eff}$, and $m_{\tilde{\chi}_1^0}/\mu_{eff} \rightarrow 1$ will increase $\sigma^{SI}$ and thus strengthen the DM-DD experimental constraints.

\begin{table}[t]
	\centering
	\caption{Signal of final state for the electroweakino pair production processes considered in this work. Relevant experimental analyses were performed using a simplified model by ATLAS and CMS collaborations, and their results have been encoded in the SmodelS-1.2.3~\cite{Khosa:2020zar}.}
	\label{tab:my-table}
	\vspace{0.3cm}
	\resizebox{1\textwidth}{!}{%
		\renewcommand\arraystretch{0.9}
		\begin{tabular}{lccr}
			\hline\hline
			\bf Name & \bf Simplified Scenario  & \bf Signal of Final State & \bf Luminosity \bf($\bm {fb^{-1}}$\bf) \\ \hline
			\multicolumn{4}{c}{\bf13 TeV} \\   \hline
			\begin{tabular}[l]{@{}l@{}}\bf CMS-SUS-17-010~\cite{Sirunyan:2018lul}\\ (arXiv:1807.07799)\end{tabular}   &\begin{tabular}[c]{@{}c@{}}$\bm{\tilde{\chi}_1^{\pm}\tilde{\chi}_1^{\mp}\rightarrow W^{\pm}\tilde{\chi}_1^0 W^{\mp}\tilde{\chi}_1^0}$\\$\bm{\tilde{\chi}_1^{\pm}\tilde{\chi}_1^{\mp}\rightarrow \nu\tilde{\ell} \ell\tilde{\nu}}$ \\ \end{tabular}&\bf2$\bm \ell$ \bf + $\bm{E_{\rm T}^{\rm miss}}$    & \bf 35.9  \\ \\
			\begin{tabular}[l]{@{}l@{}}\bf CMS-SUS-17-009~\cite{Sirunyan:2018nwe}\\ (arXiv:1806.05264)\end{tabular}   &$\bm{\tilde{\ell}\tilde{\ell}}$ &\bf2$\bm \ell$ \bf + $\bm{E_{\rm T}^{\rm miss}}$    & \bf 35.9               \\ \\
			\begin{tabular}[l]{@{}l@{}}\bf CMS-SUS-17-004~\cite{Sirunyan:2018ubx}\\ (arXiv:1801.03957)\end{tabular} &$\bm{\tilde{\chi}_{2}^0\tilde{\chi}_1^{\pm}\rightarrow Wh(Z)\tilde{\chi}_1^0\tilde{\chi}_1^0$} & \bf n$\bm \ell$\bf(n\textgreater{}=0) + \bf nj(n\textgreater{}=0) + $\bm E_{\rm T}^{\rm miss}$   & \bf 35.9               \\ \\
			\begin{tabular}[l]{@{}l@{}}\bf CMS-SUS-16-045~\cite{Sirunyan:2017eie}\\ (arXiv:1709.00384)\end{tabular}          &$\bm{\tilde{\chi}_2^0\tilde{\chi}_1^{\pm}\rightarrow W^{\pm}\tilde{\chi}_1^0h\tilde{\chi}_1^0}$& \bf 1$\bm \ell$\bf 2b + $\bm E_{\rm T}^{\rm miss}$                           & \bf 35.9               \\ \\
			\begin{tabular}[l]{@{}l@{}}\bf CMS-SUSY-16-039~\cite{Sirunyan:2017lae}\\(arxiv:1709.05406) \end{tabular}          &\begin{tabular}[c]{@{}c@{}c@{}c@{}c@{}} $\bm {\tilde{\chi}_2^0\tilde{\chi}_1^{\pm}\rightarrow \ell\tilde{\nu}\ell\tilde{\ell}}$\\$\bm{\tilde{\chi}_2^0\tilde{\chi}_1^{\pm}\rightarrow\tilde{\tau}\nu\tilde{\ell}\ell}$\\$\bm{\tilde{\chi}_2^0\tilde{\chi}_1^{\pm}\rightarrow\tilde{\tau}\nu\tilde{\tau}\tau}$\\ $\bm{\tilde{\chi}_2^0\tilde{\chi}_1^{\pm}\rightarrow WZ\tilde{\chi}_1^0\tilde{\chi}_1^0}$\\$\bm{\tilde{\chi}_2^0\tilde{\chi}_1^{\pm}\rightarrow WH\tilde{\chi}_1^0\tilde{\chi}_1^0}$\end{tabular} & \bf n$\bm{\ell(n\textgreater{}0)}$(\bm{$\tau}$) \bf + $\bm{E_{\rm T}^{\rm miss}}$                           & \bf 35.9               \\ \\
			\begin{tabular}[l]{@{}l@{}}\bf CMS-SUS-16-034~\cite{Sirunyan:2017qaj}\\ (arXiv:1709.08908)\end{tabular}&$\bm{\tilde{\chi}_2^0\tilde{\chi}_1^{\pm}\rightarrow W\tilde{\chi}_1^0Z(h)\tilde{\chi}_1^0}$ & \bf n$\bm{\ell}$\bf (n\textgreater{}=2) + nj(n\textgreater{}=1) $\bm{E_{\rm T}^{\rm miss}}$                           & \bf 35.9               \\ \\
			\begin{tabular}[l]{@{}l@{}}\bf ATLAS-1803-02762~\cite{Aaboud:2018jiw}\\ (ATLAS-SUSY-2016-24)\end{tabular} &\begin{tabular}[c]{@{}c@{}c@{}c@{}}$ \bm {\tilde{\chi}_2^0\tilde{\chi}_1^{\pm}\rightarrow WZ\tilde{\chi}_1^0\tilde{\chi}_1^0}$\\$\bm{\tilde{\chi}_2^0\tilde{\chi}_1^{\pm}\rightarrow \nu\tilde{\ell}l\tilde{\ell}}$\\$\bm {\tilde{\chi}_1^{\pm}\tilde{\chi}_1^{\mp}\rightarrow \nu\tilde{\ell}\nu\tilde{\ell}}$\\ $\bm{ \tilde{\ell}\tilde{\ell}}$\end{tabular} & \bf n$\bm \ell$\bf (n\textgreater{}=2) + $\bm E_{\rm T}^{\rm miss}$ & \bf 36.1               \\ \\
			\begin{tabular}[l]{@{}l@{}}\bf ATLAS-1812-09432~\cite{Aaboud:2018ngk}\\ (ATLAS-SUSY-2017-01)\end{tabular} &$\bm{\tilde{\chi}_2^0\tilde{\chi}_1^{\pm}\rightarrow Wh\tilde{\chi}_1^0\tilde{\chi}_1^0}$ & \bf n$\bm \ell$\bf (n\textgreater{}=0) + nj(n\textgreater{}=0) + nb(n\textgreater{}=0) + n$\bm \gamma$\bf (n\textgreater{}=0) + $\bm E_{\rm T}^{\rm miss}$ & \bf 36.1               \\ \\
			\begin{tabular}[l]{@{}l@{}}\bf ATLAS-1806-02293~\cite{Aaboud:2018sua}\\ (ATLAS-SUSY-2017-03)\end{tabular} &$\bm{\tilde{\chi}_2^0\tilde{\chi}_1^{\pm}\rightarrow WZ\tilde{\chi}_1^0\tilde{\chi}_1^0}$ & \bf n$\bm \ell$\bf (n\textgreater{}=2) + nj(n\textgreater{}=0) + $\bm E_T^{miss}$ & \bf 36.1               \\ \\
			\begin{tabular}[l]{@{}l@{}}\bf ATLAS-1912-08479~\cite{Aad:2019vvi}\\ (ATLAS-SUSY-2018-06)\end{tabular}          &$\bm{\tilde{\chi}_2^0\tilde{\chi}_1^{\pm}\rightarrow W(\rightarrow l\nu)\tilde{\chi}_1^0Z(\rightarrow\ell\ell)\tilde{\chi}_1^0}$& \bf 3$\bm \ell $ \bf + $\bm E_{\rm T}^{\rm miss}$                           & \bf 139               \\ \\
			\begin{tabular}[l]{@{}l@{}}\bf ATLAS-1908-08215~\cite{Aad:2019vnb}\\ (ATLAS-SUSY-2018-32)\end{tabular}   &\begin{tabular}[c]{@{}c@{}}$\bm{\tilde{\ell}\tilde{\ell}}$\\$\bm{\tilde{\chi}_1^{\pm}\tilde{\chi}_1^{\mp}}$ \\ \end{tabular} & \bf 2$\bm \ell$ \bf + $\bm E_{\rm T}^{\rm miss}$ & \bf 139               \\ \\
			\begin{tabular}[l]{@{}l@{}}\bf ATLAS-1909-09226~\cite{Aad:2019vvf}\\ (ATLAS-SUSY-2019-08)\end{tabular}          & $\bm{\tilde{\chi}_{2}^0\tilde{\chi}_1^{\pm}\rightarrow Wh\tilde{\chi}_1^0\tilde{\chi}_1^0}$                            & \bf 1$\bm \ell$ \bf + h\bf($\bm \rightarrow$\bf bb) + $\bm E_{\rm T}^{\rm miss}$    & \bf 139               \\ \hline
			
			\multicolumn{4}{c}{\bf 8 TeV} \\  \hline
			\begin{tabular}[l]{@{}l@{}}\bf ATLAS-CONF-2013-035~\cite{ATLAS:2013rla}\end{tabular}&\begin{tabular}[c]{@{}c@{}}$\bm{\tilde{\chi}_2^0\tilde{\chi}_1^{\pm}\rightarrow Z^{(*)}\tilde{\chi}_1^0W^{(*)}\tilde{\chi}_1^0}$\\$\bm{\tilde{\chi}_2^0\tilde{\chi}_1^{\pm}\rightarrow
				\tilde{l}(\tilde{\nu})\ell(\nu)\tilde{\ell}(\tilde{\nu})\nu(\tilde{\ell})}$ \end{tabular} & \bf n$\bm \ell $\bf (n\textgreater{}=2) + $\bm E_{\rm T}^{\rm miss}$                           & \bf 20.3         \\ \\
			\begin{tabular}[l]{@{}l@{}}\bf ATLAS-CONF-2013-049~\cite{Sirunyan:2018iwl}\\(arxiv:1801.01846) \end{tabular}          &\begin{tabular}[c]{@{}c@{}} $\bm{\tilde{\ell}\tilde{\ell}}$\\$\bm{\tilde{\chi}_1^{\pm}\tilde{\chi}_1^{\mp}\rightarrow \tilde{\ell}\nu(\tilde{\nu}\ell)}$\end{tabular} &\bf  2$\bm \ell$ \bf + $\bm E_{\rm T}^{\rm miss}$                           & \bf 20.3               \\ \\
			\begin{tabular}[l]{@{}l@{}}\bf ATLAS-CONF-2013-093~\cite{TheATLAScollaboration:2013zia} \end{tabular}          &\begin{tabular}[c]{@{}c@{}} $\bm{\tilde{\chi}_2^0\tilde{\chi}_1^{\pm}\rightarrow W\tilde{\chi}_1^0h\tilde{\chi}_1^0}$\end{tabular} & \bf 1$\bm \ell$ \bf + 2$\bm b$ \bf + $\bm E_{\rm T}^{\rm miss}$                           & \bf 20.3               \\ \\
			\begin{tabular}[l]{@{}l@{}}\bf ATLAS-1403-5294~\cite{Aad:2014vma}\\(ATLAS-SUSY-2013-11)\end{tabular}&\begin{tabular}[c]{@{}c@{}c@{}c@{}}$\bm{\tilde{\chi}_1^{\pm}\tilde{\chi}_1^{\mp}\rightarrow \tilde{l}(\tilde{\nu})\nu(l)}$\\$\bm{\tilde{\chi}_1^{\pm}\tilde{\chi}_1^{\mp}\rightarrow
				W\tilde{\chi}_1^0 W\tilde{\chi}_1^0}$\\$\bm{\tilde{\chi}_1^{\pm}\tilde{\chi}_2^{0}\rightarrow W\tilde{\chi}_1^0 Z\tilde{\chi}_1^0}$\\$\bm{\tilde{\ell}\tilde{\ell}}$ \end{tabular} & \bf n$\bm \ell$\bf (n\textgreater{}=2) + $\bm{E_{\rm T}^{\rm miss}}$                          & \bf 20.3               \\ \\
			\begin{tabular}[l]{@{}l@{}}\bf ATLAS-1402-7029~\cite{Aad:2014nua}\\ (ATLAS-SUSY-2013-12)\end{tabular}          &\begin{tabular}[c]{@{}c@{}c@{}c@{}}$\bm{\tilde{\chi}_2^0\tilde{\chi}_1^{\pm}\rightarrow W\tilde{\chi}_1^0Z\tilde{\chi}_1^0}$\\
				$\bm{\tilde{\chi}_2^0\tilde{\chi}_1^{\pm}\rightarrow W\tilde{\chi}_1^0h\tilde{\chi}_1^0}$\\
				$\bm{\tilde{\chi}_2^0\tilde{\chi}_1^{\pm}\rightarrow \ell\tilde{\nu}\ell\tilde{\ell}(\tilde{\nu}\nu)}$,$\bm {\ell\tilde{\ell}\nu\tilde{\ell}(\tilde{\nu}\nu)}$\\
				 $\bm{\tilde{\chi}_2^0\tilde{\chi}_1^{\pm}\rightarrow\tilde{\tau}\nu\tilde{\tau}\tau(\tilde{\nu}\nu)}$,$\bm{\tau\tilde{\nu}\tilde{\tau}\tau(\tilde{\nu}\nu)}$\end{tabular} & \bf 3$\bm \ell(\tau)$ \bf + $\bm E_{\rm T}^{\rm miss}$                           & \bf 20.3               \\ \\
			\begin{tabular}[l]{@{}l@{}}\bf ATLAS-1501-07110~\cite{Aad:2015jqa}\\ (ATLAS-SUSY-2013-23)\end{tabular}&$\bm{\tilde{\chi}_2^0\tilde{\chi}_1^{\pm}\rightarrow W\tilde{\chi}_1^0h\tilde{\chi}_1^0}$ & \bf n$\bm\ell$
			\bf (n$\bm \textgreater{}0$\bf ) + n$\bm \gamma$\bf (n\textgreater{}=0) + n$\bm b$\bf (n\textgreater{}=0) + $\bm E_{\rm T}^{\rm miss}$                           & \bf 20.3          \\ \\
			\begin{tabular}[l]{@{}c@{}}\bf CMS-PAS-SUSY-12-022\cite{CMS:2013bda}\\\end{tabular}          &\begin{tabular}[c]{@{}c@{}c@{}c@{}c@{}}
				$\bm{\tilde{\chi}_2^0\tilde{\chi}_1^{\pm}\rightarrow \tilde{\ell}\ell\tilde{\nu}\ell}$,$\bm{\ell\tilde{\ell}\nu\tilde{\ell}}$\\
				$\bm{\tilde{\chi}_2^0\tilde{\chi}_1^{\pm}\rightarrow Z\tilde{\chi}_1^0 W\tilde{\chi}_1^0}$\\$\bm{\tilde{\chi}_2^0\tilde{\chi}_3^0\rightarrow Z\tilde{\chi}_1^0 Z\tilde{\chi}_1^0}$\\$\bm{\tilde{\chi}_1^{\pm}\tilde{\chi}_1^{\mp}\rightarrow \ell\tilde{\nu}\nu\tilde{\ell}}$\\$\bm{\tilde{\ell}\tilde{\ell}}$\end{tabular} &\bf n$\bm \ell$\bf (n$\bm \textgreater{}=2$\bf ) + $\bm E_{\rm T}^{\rm miss}$& \bf 9.2 \\ \\
			\begin{tabular}[l]{@{}c@{}}\bf CMS-SUSY-13-006~\cite{CMS:2013dea}\\\end{tabular}          &\begin{tabular}[c]{@{}c@{}c@{}c@{}}
				$\bm{\tilde{\chi}_2^0\tilde{\chi}_1^{\pm}\rightarrow \tilde{\ell}\ell\tilde{\nu}\ell}$,$\bm{\ell\tilde{\ell}\nu\tilde{\ell}}$\\
				$\bm{\tilde{\chi}_2^0\tilde{\chi}_1^{\pm}\rightarrow Z\tilde{\chi}_1^0 W\tilde{\chi}_1^0}$\\$\bm{\tilde{\chi}_1^{\pm}\tilde{\chi}_1^{\mp}\rightarrow \ell\tilde{\nu}\nu\tilde{\ell}}$\\$\bm{\tilde{\ell}\tilde{\ell}}$\end{tabular} & \bf n$\bm \ell$\bf (n\textgreater{}=2) + $\bm E_{\rm T}^{\rm miss}$& \bf 19.5   \\       \hline\hline
			
	\end{tabular}} 
\end{table}

\section{\label{sec:scan} Model scans and constraints}
The \texttt{NMSSMTools-5.4.1} package~\cite{Ellwanger:2004xm,Ellwanger:2005dv} was used to produce samples of singlino-dominated DM scenarios in the $\rm Z_3$-NMSSM and to model the corresponding features in detail\footnote{The \texttt{NMSSMTools} package includes codes to compute various observables in Higgs physics, DM physics, B physics, and sparticle physics. In this sector, we only briefly introduce the calculation of the observables we are interested in.}. A sophisticated scan was first performed over the following ranges in the parameter space:
\begin{eqnarray}\label{parameter-scan}
&&  0 < \lambda \leq 0.7,~~|\kappa| \leq 0.7,~~1 \leq \tan \beta \leq 60,~~100 ~{\rm GeV} \leq \mu_{eff} \leq 1000~{\rm GeV}, \nonumber
\\
&& |A_\kappa| < 1 ~{\rm TeV},~~0 <A_\lambda \leq 5~{\rm TeV}, ~~|A_t| \leq 5~{\rm TeV}, ~~|M_1| \leq 500~{\rm GeV},  \label{Parameters}
\end{eqnarray}
in which all parameters were defined at the scale $Q = 1~{\rm TeV}$. Upper bounds of $0.7$ were imposed on $\lambda$ and $|\kappa|$ to maintain a perturbable theory up to the grand unification scale. A lower bound of $100~{\rm GeV}$ was placed on $\mu_{eff}$ by the LEP search for electroweakinos~\cite{Tanabashi:2018oca}, and an upper bound of 1000 GeV for $\mu_{eff}$ is large enough to allow us to consider various possibilities (see the discussion presented below). In addition, noting that the LHC search for SUSY prefers massive charged sparticles, the following assumptions were made concerning unimportant SUSY parameters. The electroweak gaugino masses were set to $M_2=2~{\rm TeV}$, and the gluino masses were set to $M_3=5~{\rm TeV}$. Soft SUSY-breaking parameters in the squarks sector were fixed at $2~{\rm TeV}$, excluding trilinear couplings $A_t = A_b$ used as free parameters to adjust the Higgs mass spectrum to coincide with relevant experimental measurements at the LHC. In addition, all slepton soft parameters were set to $2~{\rm TeV}$, as we did not want to explain the muon $g$-2 anomaly. We also required $\lambda \geq 2 |\kappa|$ in the scan to achieve a singlino-dominated $\tilde\chi^0_1$.

Specifically, the MultiNest algorithm~\cite{Feroz:2008xx,Feroz:2013hea} with flat distributions for all the parameters in Eq.~(\ref{Parameters}) and {\it nlive} = 20000 were adopted during the scan to ensure that the conclusions were as complete as possible, and more than 20 thousands CPU hours were spent on the calculations\footnote{The multi-nest sampling algorithm explores a high-dimensional parameter space by determining the iso-likelihood contour in each iteration with {\it nlive} active points (the integer {\it nlive} is an input parameter of the algorithm, and it usually takes a value larger than 1000. In general, the larger value {\it nlive} adopts, the more accurate the scan result becomes.)~\cite{Feroz:2008xx,Feroz:2013hea}. It is good at dealing with the case in which the samples' posterior distribution is multi-modal or degenerate, which is frequently encountered in new physics studies. In contrast, the Markov Chain method~\cite{Markov-Chain} is highly inefficient for such a situation, and thus, it usually provides incomplete information about the distribution.}. Several constraints were imposed by constructing the following corresponding likelihood function to guide the process:
\begin{eqnarray}
\mathcal{L}=\mathcal{L}_{\rm{m_{h}}} \times \mathcal{L}_{h, {\rm coupling}} \times {\mathcal{L}_{B} } \times  {\mathcal{L}_{EW} } \times  {\mathcal{L}_{\Omega h^2} }\times  {\mathcal{L}_{DD}},
\label{Likelihood}
\end{eqnarray}
where $\mathcal{L}_{\rm{m_{h}}}$ and $\mathcal{L}_{h, {\rm coupling}}$ are likelihood functions for the experimentally measured SM-like Higgs boson mass and couplings, respectively.
The computation of $m_h$ included leading electroweak corrections, two loop terms, and propagator corrections, as in Ref. \cite{Degrassi:2009yq}. Its experimental central value was taken as $m_h = 125.09~{\rm GeV}$~\cite{Aad:2015zhl}, and a total experimental and theoretical uncertainty of $3~{\rm GeV}$ was assumed. $\mathcal{L}_{h, {\rm coupling}}$ works in a seven-parameter $\kappa$-framework with related experimental measurements, such as the central values and uncertainties of the Higgs couplings and their correlation coefficients, taken from the ATLAS analysis, using 80${\rm fb}^{-1}$ data collected during the LHC Run-II~\cite{Aad:2019mbh}. Some knowledge about probability and statistics was used in constructing $\mathcal{L}_{h, {\rm coupling}}$ (see the introduction in Ref. \cite{Tanabashi:2018oca}). ${\mathcal{L}_{B} }$ is the likelihood function for the measurement of the branching ratio for the decays $B \to X_s \gamma$ and $B_s \to \mu^+\mu^-$. These ratios were calculated by the formulae in Refs.~\cite{Domingo:2007dx,Domingo:2015wyn}, and their experimental values were taken from Ref. \cite{Tanabashi:2018oca}. ${\mathcal{L}_{\Omega h^2} }$ and ${\mathcal{L}_{DD}}$ are likelihood functions for the measured abundance from the WMAP/Planck experiments~\cite{Ade:2015xua, Aghanim:2018eyx} and the detection of both spin-independent (SI) and spin-dependent (SD) DM-nucleon scattering in the XENON-1T experiment~\cite{Aprile:2018dbl,Aprile:2019dbj}. Relevant quantities
were calculated using the \texttt{micrOMEGAs} package~\cite{Belanger:2004yn,Belanger:2005kh,Belanger:2006is,Belanger:2013oya,Belanger:2018ccd}.
In addition, $ {\mathcal{L}_{EW} }$ denotes a likelihood function for precision electroweak observables of $\epsilon_i$ (i=1,2,3)~\cite{Altarelli:1990zd,Altarelli:1991fk,Altarelli:1994iz} or, equivalently, $S$, $T$, and $U$ parameters~\cite{Peskin:1990zt,Peskin:1991sw} calculated using the formulas from Cao and Yang~\cite{Cao:2008rc} and fitted to corresponding measurements by the procedure presented in Ref. \cite{deBlas:2016ojx}. Each of these likelihood functions was assumed to follow a Gaussian distribution, with explicit representations provided in Cao et al.~\cite{Cao:2018iyk}.

The acquired samples were further refined using the following criteria: the SM-like Higgs mass was within the range of 122--128 GeV, the observed DM relic abundance was within $\pm 10\%$ of the measured central value $\Omega h^2 = 0.1187$~\cite{Aghanim:2018eyx}, the upper bound was a $90\%$ confidence level of the XENON-1T experimental results of the SI cross section~\cite{Aprile:2018dbl} and the SD cross section~\cite{Aprile:2019dbj}, and all other constraints implemented in \texttt{NMSSMTools}, including various B-physics observables in corresponding experimentally allowed ranges, were at the $2\sigma$ level. We also required the samples to satisfy $\chi^2_{EW} \leq 7.8$ and $\chi^2_{h, {\rm coupling}} \leq 14.1$, which corresponded to $95\%$ confidence level exclusion limits for three and seven degrees of freedom, respectively. Since a more significant deviation of the electroweak precision observables (the Higgs couplings) from their measured values would enhance $\chi^2_{EW}$ ($\chi^2_{h, {\rm coupling}}$), this requirement delineated the experimentally allowed range of these observables to further limit the $Z_3$-NMSSM.

Constraints were also implemented from the LHC search for electroweakinos using the \texttt{SModelS-1.2.3} code~\cite{Ambrogi:2017neo, Kraml:2013mwa,Khosa:2020zar}. The final states listed in Table~\ref{tab:my-table} from all the electroweakino pair production processes were considered during this process\footnote{The concrete procedure to determine the limitation is as follows: we first determined the signal region (SR) with the largest expected sensitivity for a given sample, then we checked its $R$ value defined by $R \equiv S/S_{95}^{OBS}$, where $S$ stands for the number of signal events in the SR with the
statistical uncertainty considered, and $S_{95}^{OBS}$ denotes the observed limit at 95\% confidence level for the SR.  Evidently, $R$ represents the capability
of the LHC in exploring a point. $R > 1$ implies that the point is excluded; otherwise, it is allowed.}.  It is worth noting that a large portion of the samples satisfying these constraints were characterized by a small mass splitting between $\tilde{\chi}_1^0$ and $\tilde{\chi}_2^0$. However, the latest ATLAS analysis of the compressed mass spectra, acquired by searching soft di-lepton signals~\cite{Aad:2019qnd}, was not included in the \texttt{SModelS-1.2.3} code. As such, the analysis constraints are validated in the Appendix, and they were applied to each sample by elaborate Monte Carlo simulations. We verified that they were very effective in excluding the co-annihilation case. We illustrated this point in our recent publication~\cite{cao:2021Gnmssm}.

We add that we did not consider constraints from indirect DM detection experiments (i.e., the Fermi-LAT search for DM annihilation from dwarf spheroidal galaxies) as they become loose for $m_{\tilde \chi^0_1} > 100~\rm GeV $~\cite{Ackermann:2015zua}. In addition, in the co-annihilation case encountered in this work, the annihilation rate of singlino-dominated DM in present day is very small, which weakens the constraints.

\begin{table}[t]
\centering
\caption{Benchmark points satisfying various experimental constraints. Mass parameters are in units of GeV, and the DM-nucleon scattering cross sections are in units of ${\rm cm^2}$. The number preceding each annihilation process represents its fraction of contributions to the total DM annihilation cross section at the freeze-out temperature. Dots include the information of the decay modes with smaller branching ratios. } \label{table2}
\vspace{0.2cm}
\scalebox{0.7}{
\begin{tabular}{lcclcccccc}
\hline\hline
\multicolumn{5}{c|}{SM-like Higgs: $h_1$ }   & \multicolumn{5}{c}{SM-like Higgs: $h_2$}    \\ \hline
    \multicolumn{1}{c}{} & \multicolumn{2}{c}{$P_1$} & \multicolumn{2}{c|}{$P_2$} & & \multicolumn{2}{c}{$P_3$} & \multicolumn{2}{c}{$P_4$}    \\ \hline
    \multicolumn{1}{l}{$\lambda$} & \multicolumn{2}{c}{0.696} & \multicolumn{2}{c|}{0.028} &
    & \multicolumn{2}{c}{0.108} & \multicolumn{2}{c}{$9.64\times10^{-3}$} \\
    \multicolumn{1}{l}{$\kappa$} & \multicolumn{2}{c}{0.208} & \multicolumn{2}{c|}{$-$0.013} &
    & \multicolumn{2}{c}{0.048} & \multicolumn{2}{c}{$-$$4.2\times10^{-3}$}
    \\
    \multicolumn{1}{l}{$\tan\beta$} & \multicolumn{2}{c}{2.15} & \multicolumn{2}{c|}{7.2} &
    & \multicolumn{2}{c}{12.8} & \multicolumn{2}{c}{9.9}
    \\
    \multicolumn{1}{l}{$\mu_{eff}$} & \multicolumn{2}{c}{683.7} & \multicolumn{2}{c|}{227.8} &
    & \multicolumn{2}{c}{206.1} & \multicolumn{2}{c}{164.4}
    \\
    \multicolumn{1}{l}{$A_t$} & \multicolumn{2}{c}{$-$2734} & \multicolumn{2}{c|}{2754} &
    & \multicolumn{2}{c}{$-$3394} & \multicolumn{2}{c}{$-$3510}
    \\
    \multicolumn{1}{l}{$A_\lambda$} & \multicolumn{2}{c}{1128} & \multicolumn{2}{c|}{3908} &
    & \multicolumn{2}{c}{3125} & \multicolumn{2}{c}{1142}
    \\
    \multicolumn{1}{l}{$A_\kappa$} & \multicolumn{2}{c}{$-$28.3} & \multicolumn{2}{c|}{24.4} &
    & \multicolumn{2}{c}{$-$323.1} & \multicolumn{2}{c}{147.3}
    \\
    \multicolumn{1}{l}{$M_1$} & \multicolumn{2}{c}{$-$471.3} & \multicolumn{2}{c|}{$-$418.2} &
     & \multicolumn{2}{c}{$-$484.7} & \multicolumn{2}{c}{211.8}
    \\
    \hline
    \multicolumn{1}{l}{$m_{\tilde{\chi}^0_1}$} & \multicolumn{2}{c}{422.4} & \multicolumn{2}{c|}{$-$216.5} &
     & \multicolumn{2}{c}{181.9} & \multicolumn{2}{c}{$-$148.6}
     \\
     \multicolumn{1}{l}{$m_{\tilde {\chi}^0_2}$} & \multicolumn{2}{c}{$-$473.7} & \multicolumn{2}{c|}{$-$233.0} &
    & \multicolumn{2}{c}{213.2} & \multicolumn{2}{c}{151.9}
    \\
    \multicolumn{1}{l}{$m_{\tilde {\chi}^0_3}$} & \multicolumn{2}{c}{702.3} & \multicolumn{2}{c|}{234.5} &
    & \multicolumn{2}{c}{$-$215.1} & \multicolumn{2}{c}{$-$174.5}
    \\
    \multicolumn{1}{l}{$m_{\tilde {\chi}^\pm_1}$} & \multicolumn{2}{c}{694.9} & \multicolumn{2}{c|}{233.2} &
    & \multicolumn{2}{c}{211.4} & \multicolumn{2}{c}{170.4}
    \\
    \multicolumn{1}{l}{$N_{13}$} & \multicolumn{2}{c}{$-$0.035} & \multicolumn{2}{c|}{$-$0.164} &
    & \multicolumn{2}{c}{$-$0.32} & \multicolumn{2}{c}{$-$0.032}
    \\
    \multicolumn{1}{l}{$N_{14}$} & \multicolumn{2}{c}{0.176} & \multicolumn{2}{c|}{$-$0.171} &
    & \multicolumn{2}{c}{$-$0.35} & \multicolumn{2}{c}{$-$0.038}
    \\
    \multicolumn{1}{l}{$N_{15}$} & \multicolumn{2}{c}{0.984} & \multicolumn{2}{c|}{0.971} &
    & \multicolumn{2}{c}{0.87} & \multicolumn{2}{c}{0.999}
    \\
    \multicolumn{1}{l}{$Z_h$} & \multicolumn{2}{c}{0.032} & \multicolumn{2}{c|}{0.056} &
    & \multicolumn{2}{c}{0.23} & \multicolumn{2}{c}{0.003}
    \\
    \multicolumn{1}{l}{$Z_S$} & \multicolumn{2}{c}{0.968} & \multicolumn{2}{c|}{0.943} &
    & \multicolumn{2}{c}{0.76} & \multicolumn{2}{c}{0.997}
    \\
    \hline
    \multicolumn{1}{l}{$m_{h_1}$} & \multicolumn{2}{c}{124.6} & \multicolumn{2}{c|}{125.2} &
    & \multicolumn{2}{c}{64.2} & \multicolumn{2}{c}{100.7}
    \\
    \multicolumn{1}{l}{$m_{h_2}$} & \multicolumn{2}{c}{440.4} & \multicolumn{2}{c|}{206.5} &
    & \multicolumn{2}{c}{126.7} & \multicolumn{2}{c}{125.3}
    \\
    \multicolumn{1}{l}{$m_{h_3}$} & \multicolumn{2}{c}{1536} & \multicolumn{2}{c|}{2532} &
    & \multicolumn{2}{c}{2905} & \multicolumn{2}{c}{1311}
    \\
    \multicolumn{1}{l}{$m_{a_1}$} & \multicolumn{2}{c}{178.0} & \multicolumn{2}{c|}{88.1} &
    & \multicolumn{2}{c}{298.0} & \multicolumn{2}{c}{178.4}
    \\
    \multicolumn{1}{l}{$m_{a_2}$} & \multicolumn{2}{c}{1536} & \multicolumn{2}{c|}{2532} &
    & \multicolumn{2}{c}{2905} & \multicolumn{2}{c}{1311}
    \\
    \multicolumn{1}{l}{$V_{h_1}^{\rm NSM}, V_{h_2}^{\rm NSM}$} & \multicolumn{2}{c}{0.0, $-$0.05} & \multicolumn{2}{c|}{$-$0.0, $-$0.0} &
    & \multicolumn{2}{c}{$-$0.0, $-$0.0} & \multicolumn{2}{c}{$-$0.0, $-$0.0}
    \\
    \multicolumn{1}{l}{$V_{h_1}^{\rm SM}, V_{h_2}^{\rm SM}$} & \multicolumn{2}{c}{0.99, 0.15} & \multicolumn{2}{c|}{0.99, $-$0.11} &
    & \multicolumn{2}{c}{0.147, 0.989} & \multicolumn{2}{c}{0.04, 0.999}
    \\
    \multicolumn{1}{l}{$V_{h_1}^{\rm S}, V_{h_2}^{\rm S}$} & \multicolumn{2}{c}{$-$0.15, 0.99} & \multicolumn{2}{c|}{0.11, 0.99} &
    & \multicolumn{2}{c}{0.989, $-$0.147} & \multicolumn{2}{c}{0.999, $-$0.04}
    \\
    \multicolumn{1}{l}{$V_{a_1}^{' \rm NSM}, V_{a_1}^{' \rm S}$} & \multicolumn{2}{c}{$-$0.03, 1.0} & \multicolumn{2}{c|}{$-$0.0, 1.0} &
    & \multicolumn{2}{c}{$-$0.0, 1.0} & \multicolumn{2}{c}{$-$0.0, 1.0}
    \\
    \hline
    \multicolumn{1}{l}{} & \multicolumn{1}{c}{} & \multicolumn{1}{c}{} & \multicolumn{1}{c}{} & \multicolumn{1}{c|}{} & \multicolumn{1}{c}{} & \multicolumn{1}{c}{} & \multicolumn{1}{c}{} & \multicolumn{1}{c}{}
    \\
    \multicolumn{1}{l}{$\sigma^{SI}_{\tilde\chi^0_1-p}$} & \multicolumn{2}{c}{$4.2 \times 10^{-49}$} & \multicolumn{2}{c|}{$3.1\times 10^{-47}$} &
    & \multicolumn{2}{c}{$8.89 \times 10^{-47}$} & \multicolumn{2}{c}{$2.67 \times 10^{-49}$}
    \\
    \multicolumn{1}{l}{$\sigma^{SD}_{\tilde\chi^0_1-n}$} & \multicolumn{2}{c}{$3.0 \times 10^{-41}$} & \multicolumn{2}{c|}{$2.94 \times 10^{-43}$} &
    & \multicolumn{2}{c}{$2.23 \times 10^{-41}$} & \multicolumn{2}{c}{$6.57 \times 10^{-45}$}
    \\
     \multicolumn{1}{l}{$\Omega h^2$} & \multicolumn{2}{c}{0.118} & \multicolumn{2}{c|}{0.109}
    & & \multicolumn{2}{c}{0.12} & \multicolumn{2}{c}{0.108}
    \\
    \multicolumn{1}{l}{} & \multicolumn{1}{c}{} & \multicolumn{1}{c}{} & \multicolumn{1}{c}{} & \multicolumn{1}{c|}{} & \multicolumn{1}{c}{} & \multicolumn{1}{c}{} & \multicolumn{1}{c}{}
    \\
    \multicolumn{1}{c}{} & \multicolumn{1}{l}{$68.4\%$} & \multicolumn{1}{l}{$\tilde\chi^0_1 \tilde\chi^0_1 \rightarrow t \bar{t}$} & \multicolumn{1}{l}{$14.0\%$} & \multicolumn{1}{l|}{$\tilde\chi^0_2 \tilde\chi^+_1 \rightarrow u \bar{d}$}
    & \multicolumn{1}{l}{} & \multicolumn{1}{l}{$28.7\%$} & \multicolumn{1}{l}{$\tilde\chi^0_1 \tilde\chi^0_1 \rightarrow W^+ W^-$}& \multicolumn{1}{l}{$35.0\%$} & \multicolumn{1}{l}{$\tilde\chi^0_2 \tilde\chi^0_2 \rightarrow W^+ W^-$}
    \\
    \multicolumn{1}{c}{}  & \multicolumn{1}{l}{$19.4\%$} & \multicolumn{1}{l}{$~~~~~~\rightarrow h_s a_s$} & \multicolumn{1}{l}{$5.4\%$} & \multicolumn{1}{l|}{$~~~~~~\rightarrow \nu_l \bar{l}$}
    & \multicolumn{1}{l}{} & \multicolumn{1}{l}{$22.0\%$}
    & \multicolumn{1}{l}{$~~~~~~\rightarrow Z Z$}& \multicolumn{1}{l}{$25.4\%$} & \multicolumn{1}{l}{$~~~~~~\rightarrow Z Z$}
    \\
    \multicolumn{1}{c}{} & \multicolumn{1}{l}{$5.0\%$} & \multicolumn{1}{l}{$~~~~~~\rightarrow h_{sm} a_s$} & \multicolumn{1}{l}{$13.0\%$} & \multicolumn{1}{l|}{$\tilde\chi^0_3 \tilde\chi^+_1 \rightarrow u \bar{d}$}
    & \multicolumn{1}{l}{} & \multicolumn{1}{l}{$17.6\%$}
    & \multicolumn{1}{l}{$~~~~~~\rightarrow t \bar{t}$} & \multicolumn{1}{l}{$16.2\%$} & \multicolumn{1}{l}{$\tilde\chi^0_2 \tilde\chi^+_1 \rightarrow u \bar{d}$}
    \\
    \multicolumn{1}{c}{} & \multicolumn{1}{l}{$2.2\%$} & \multicolumn{1}{l}{$~~~~~~ \rightarrow W^+ W^-$} & \multicolumn{1}{l}{$4.8\%$} & \multicolumn{1}{l|}{$~~~~~~\rightarrow \nu_l \bar{l}$} & \multicolumn{1}{c}{} & \multicolumn{1}{l}{$13.0\%$} & \multicolumn{1}{l}{$\tilde\chi^0_1 \tilde\chi^+_1 \rightarrow u \bar{d}$}& \multicolumn{1}{l}{$6.6\%$} & \multicolumn{1}{l}{$~~~~~~\rightarrow \nu_l \bar{l}$}
    \\
    \multicolumn{1}{l}{annihilation} & \multicolumn{1}{l}{...} & \multicolumn{1}{l}{~~~~~...} & \multicolumn{1}{l}{$11.4\%$} & \multicolumn{1}{l|}{$\tilde\chi^0_1 \tilde\chi^+_1 \rightarrow u \bar{d}$} & & \multicolumn{1}{c}{$4.8\%$} & \multicolumn{1}{l}{$~~~~~~\rightarrow \nu_l \bar{l}$}& \multicolumn{1}{l}{4.1\%} & \multicolumn{1}{l}{$\tilde\chi^0_2 \tilde\chi^0_3 \rightarrow q \bar{q}$}
    \\
    \multicolumn{1}{l}{channels} & \multicolumn{1}{c}{} & \multicolumn{1}{l}{} & \multicolumn{1}{l}{4.2\%} & \multicolumn{1}{l|}{$~~~~~~\rightarrow \nu_l \bar{l}$}
    & & \multicolumn{1}{l}{$3.8\%$} & \multicolumn{1}{l}{$\tilde\chi^0_1 \tilde\chi^0_3 \rightarrow q \bar{q}$}& \multicolumn{1}{l}{1.2\%} & \multicolumn{1}{l}{$~~~~~~\rightarrow \nu_l \bar{\nu}_l$}
    \\
    \multicolumn{1}{l}{}                  & \multicolumn{1}{l}{} & \multicolumn{1}{l}{} & \multicolumn{1}{l}{6.8\%} & \multicolumn{1}{l|}{$\tilde\chi^0_2 \tilde\chi^0_3 \rightarrow q \bar{q}$} & \multicolumn{1}{l}{}& \multicolumn{1}{l}{$...$} & \multicolumn{1}{l}{$~~~~~~...$}& \multicolumn{1}{l}{3.6\%} & \multicolumn{1}{l}{$\tilde\chi^0_1 \tilde\chi^0_2 \rightarrow q \bar{q}$}
    \\
    \multicolumn{1}{l}{}                  & \multicolumn{1}{l}{} & \multicolumn{1}{l}{} & \multicolumn{1}{l}{1.8\%} & \multicolumn{1}{l|}{$~~~~~~ \rightarrow \nu_l \bar{\nu}_l$}
     & \multicolumn{1}{l}{}& \multicolumn{1}{l}{} & \multicolumn{1}{l}{}& \multicolumn{1}{l}{...} & \multicolumn{1}{l}{~~~~~...}
    \\
    \multicolumn{1}{l}{}                  & \multicolumn{1}{l}{} & \multicolumn{1}{l}{} & \multicolumn{1}{l}{7.8\%} & \multicolumn{1}{l|}{$\tilde\chi^+_1 \tilde\chi^-_1 \rightarrow q \bar{q}$} & \multicolumn{1}{l}{}        & \multicolumn{1}{l}{} & \multicolumn{1}{l}{} & \multicolumn{1}{l}{} & \multicolumn{1}{l}{}
    \\
    \multicolumn{1}{l}{}                  & \multicolumn{1}{l}{} & \multicolumn{1}{l}{} & \multicolumn{1}{l}{2.7\%} & \multicolumn{1}{l|}{$~~~~~~\rightarrow l \bar{l}$} & \multicolumn{1}{l}{}                  & \multicolumn{1}{l}{} & \multicolumn{1}{l}{}& \multicolumn{1}{l}{} & \multicolumn{1}{l}{}
    \\
    \multicolumn{1}{l}{}                  & \multicolumn{1}{l}{} & \multicolumn{1}{l}{} & \multicolumn{1}{l}{2.2\%} & \multicolumn{1}{l|}{~~~~~~$\rightarrow W^+ W^-$}
    & \multicolumn{1}{l}{}  & \multicolumn{1}{l}{} & \multicolumn{1}{l}{}& \multicolumn{1}{c}{} & \multicolumn{1}{l}{}
    \\
    \multicolumn{1}{l}{}                  & \multicolumn{1}{l}{} & \multicolumn{1}{l}{} & \multicolumn{1}{l}{5.7\%} & \multicolumn{1}{l|}{$\tilde\chi^0_1 \tilde\chi^0_3\rightarrow q \bar{q}$}
    & \multicolumn{1}{l}{}  & \multicolumn{1}{c}{} & \multicolumn{1}{l}{} & \multicolumn{1}{l}{} & \multicolumn{1}{l}{}
    \\
    \multicolumn{1}{l}{}                  & \multicolumn{1}{l}{} & \multicolumn{1}{l}{} & \multicolumn{1}{l}{1.5\%} & \multicolumn{1}{l|}{~~~~~~$\rightarrow \nu_l \bar{\nu}_l$}
    & \multicolumn{1}{l}{}    & \multicolumn{1}{c}{} & \multicolumn{1}{l}{} & \multicolumn{1}{l}{} & \multicolumn{1}{l}{}
    \\
    \multicolumn{1}{l}{}                  & \multicolumn{1}{l}{} & \multicolumn{1}{l}{} & \multicolumn{1}{l}{~~...} & \multicolumn{1}{l|}{$~~~~~~...$}
    & \multicolumn{1}{l}{}   & \multicolumn{1}{c}{ } & \multicolumn{1}{l}{ } & \multicolumn{1}{l}{} & \multicolumn{1}{l}{}
    \\
    \multicolumn{1}{l}{}                  & \multicolumn{1}{l}{} & \multicolumn{1}{l}{} & \multicolumn{1}{c}{} & \multicolumn{1}{l|}{}
    & \multicolumn{1}{l}{}    & \multicolumn{1}{c}{} & \multicolumn{1}{l}{} & \multicolumn{1}{l}{} & \multicolumn{1}{l}{}
    \\
    \hline\hline
    \end{tabular}}
\end{table}

\begin{figure*}[t]
		\centering
		\resizebox{1.\textwidth}{!}{
        \includegraphics[width=0.90\textwidth]{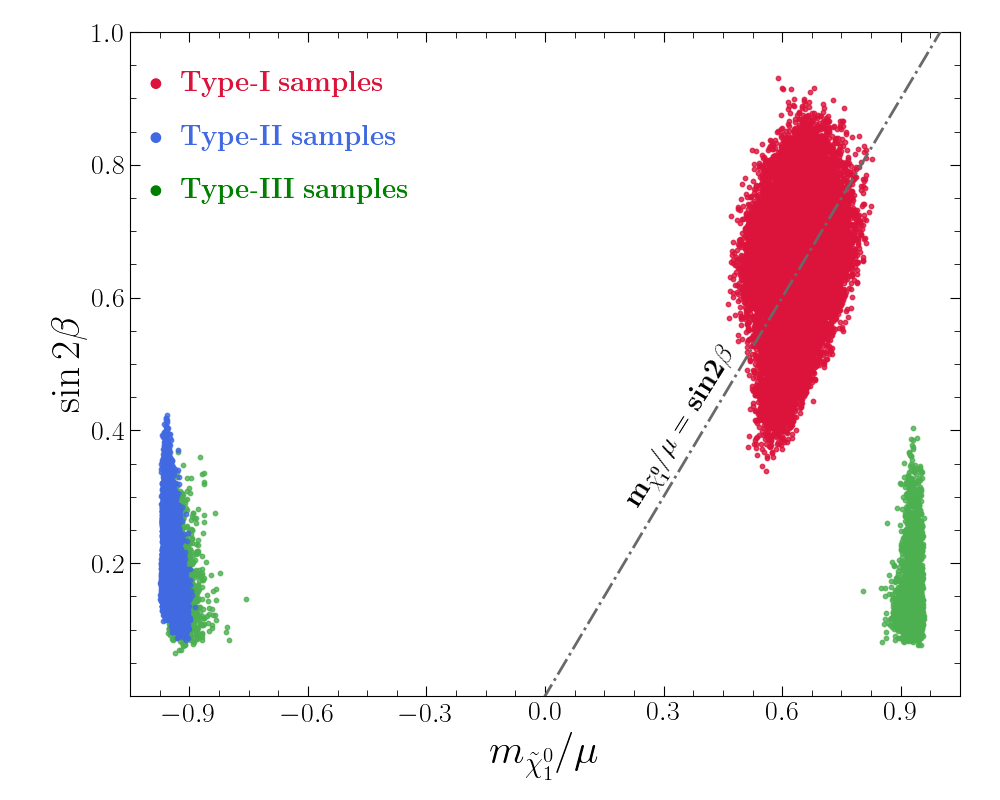}
        \includegraphics[width=0.90\textwidth]{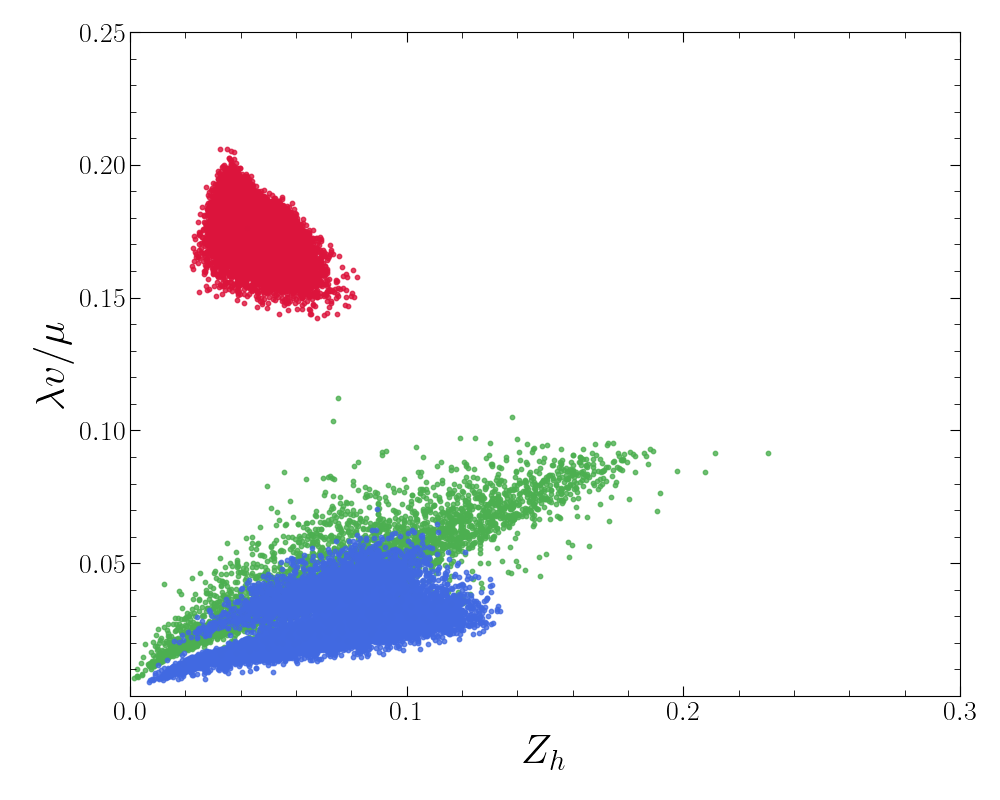}
        }
       \vspace{-0.6cm}
        \caption{Type-I, -II, and -III samples projected on the $\sin2\beta-m_{\tilde{\chi}_1^0}/\mu_{eff}$ and $\lambda v / \mu_{eff}-Z_h$ planes. \label{fig1} }
\end{figure*}	
\begin{figure*}[t]
		\centering
		\resizebox{1.03\textwidth}{!}{
        \includegraphics[width=0.90\textwidth]{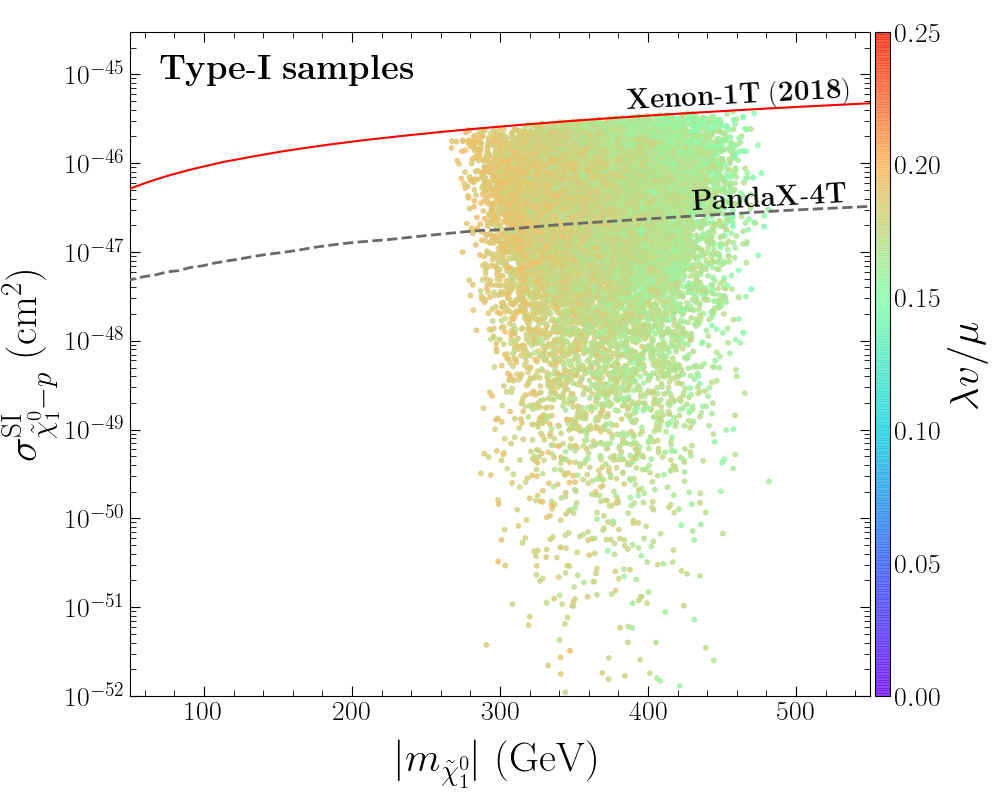}
        \includegraphics[width=0.90\textwidth]{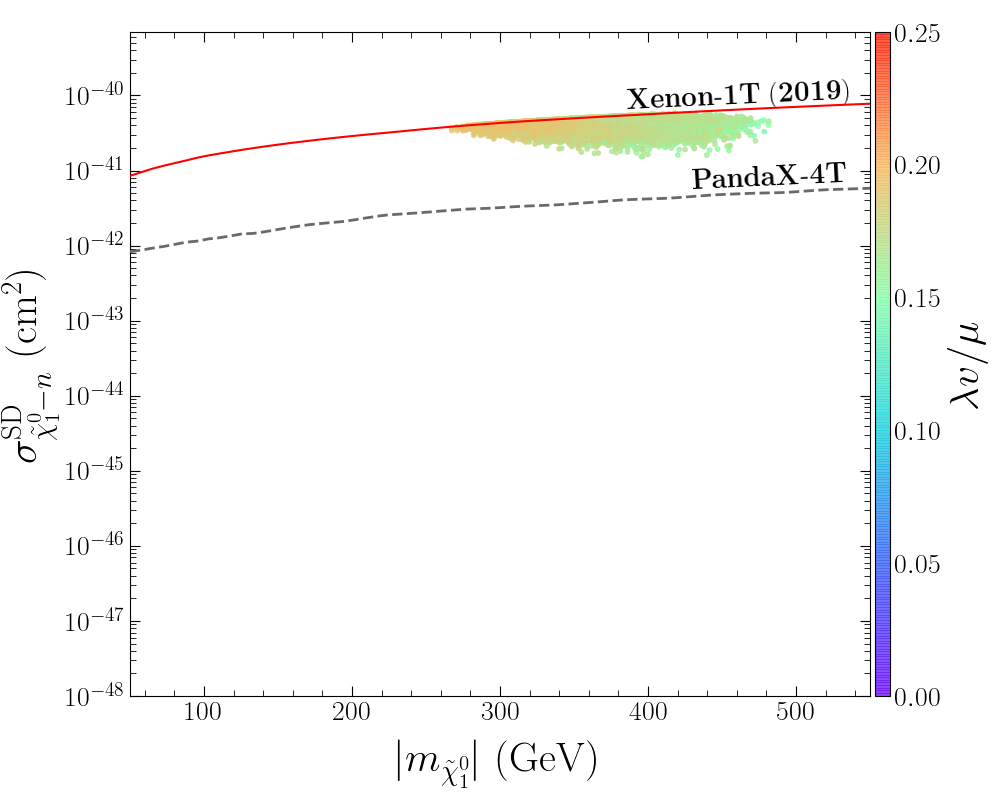}
        }
       \vspace{-0.6cm}
        \caption{Type-I samples projected on the $\sigma_{\tilde{\chi}_1^0-p}^{SI}-m_{\tilde{\chi}_1^0}$ and $\sigma_{\tilde{\chi}_1^0-n}^{SD}-m_{\tilde{\chi}_1^0}$ planes. The color bar represents the values of $\lambda v /\mu_{eff}$. The solid line represents the current exclusion bound of the XENON-1T experiment on the cross section and the dashed line denotes the projected sensitivity of the near-future PandaX-4T experiment. \label{fig2} }
\end{figure*}	

\section{\label{result13}Numerical Results}
A closer analysis suggested that the eventual surviving parameter space could be classified into the following three types:
\begin{itemize}
\item Type-I samples: the lightest $\rm {CP}$-even Higgs boson $h_1$  as the SM-like Higgs boson, $0.4\lesssim\lambda\lesssim0.7$,  $0.13 \lesssim \kappa \lesssim 0.23$, $1.5 \lesssim \tan\beta \lesssim 6 $, $450~{\rm GeV} \lesssim \mu_{eff} \lesssim 720~{\rm GeV}$, and the Bayesian evidence $\ln Z_1 = -24.2$. Due to the limited capability of the Markov Chain algorithm, this type of samples were neglected in~\cite{Cao:2018rix}.
\item Type-II samples: the lightest $\rm {CP}$-even Higgs boson $h_1$ as the SM-like Higgs boson, $\lambda\lesssim0.08$, $-0.04 \lesssim \kappa <0$ with $2 |\kappa|/\lambda \simeq 1$, $4 \lesssim \tan\beta \lesssim 24$, $170~{\rm GeV} \lesssim \mu_{eff} \lesssim 420~{\rm GeV}$, and the Bayesian evidence $\ln Z_2 = -27.5$.
\item Type-III samples: the second lightest $\rm {CP}$-even Higgs boson $h_2$ as the SM-like Higgs boson, $\lambda \lesssim 0.15$, $|\kappa| \lesssim 0.06$ with $2 |\kappa|/\lambda \simeq 1$, $ 4.5 \lesssim \tan\beta \lesssim 32$, $135~{\rm GeV} \lesssim \mu_{eff} \lesssim 260~{\rm GeV}$, and the Bayesian evidence $\ln Z_3 = -27.0$.
\end{itemize}
It is evident that the three types of parameter spaces are extremely narrow. The Bayesian evidence $Z_i$ ($i=1,2,3$) produced the Jeffreys’ scales~\cite{Bayes,Jeffreys} $\delta_{12} \equiv \ln Z_1 - \ln Z_2 = 3.3$ and $\delta_{13} \equiv \ln Z_1 - \ln Z_3 = 2.8$. These results suggest that the considered experiments slightly preferred Type-I samples to Type-II and -III samples. However, as discussed below, Type-I samples will be examined by the near-future PandaX-4T experiment~\cite{Zhang:2018xdp}. Furthermore, Type-II and -III samples share the following features in the parameter space: both $\lambda$ and $|\kappa|$ are small, $2|\kappa|/\lambda \simeq 1$, and the higgsinos are moderately light. These facts lead to the similarities of the Type-II and -III samples in many aspects of DM physics (see the discussions below for detailed similarities and differences). Nevertheless, they are still distinguished from each other in Higgs physics~\cite{Ellwanger:2011aa, Gunion:2012zd,King:2012tr, King:2012is,Cao:2012fz,Vasquez:2012hn}.

\begin{figure*}[t]
		\centering
		\resizebox{1.\textwidth}{!}{
        \includegraphics[width=0.90\textwidth]{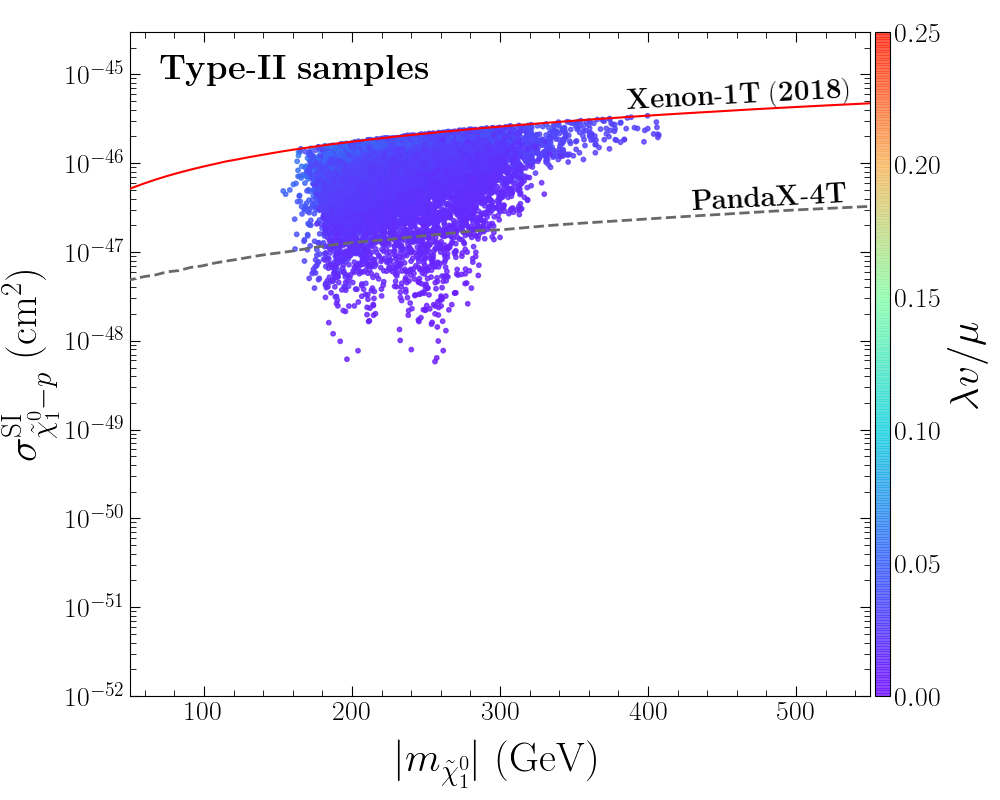}
        \includegraphics[width=0.90\textwidth]{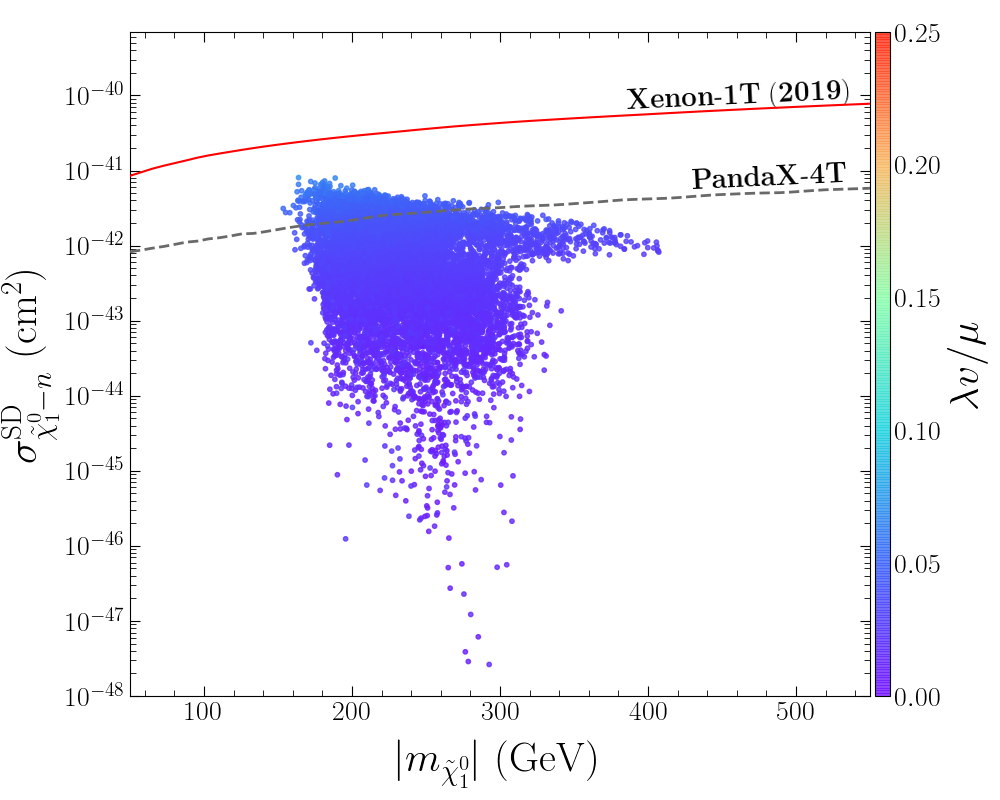}
        }
       \vspace{-0.6cm}
        \caption{Same plots as in Fig.\ref{fig2}, but for the results of the Type-II samples. \label{fig3}  }
\end{figure*}

\begin{figure*}[t]
		\centering
		\resizebox{1.\textwidth}{!}{
        \includegraphics[width=0.90\textwidth]{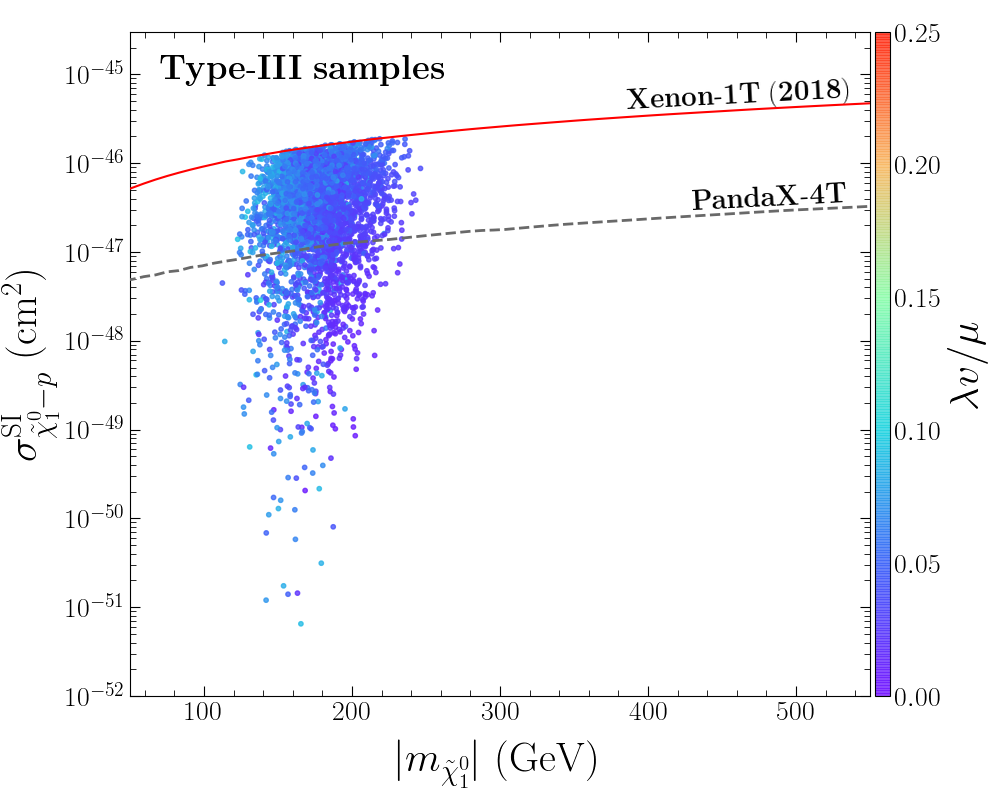}
        \includegraphics[width=0.90\textwidth]{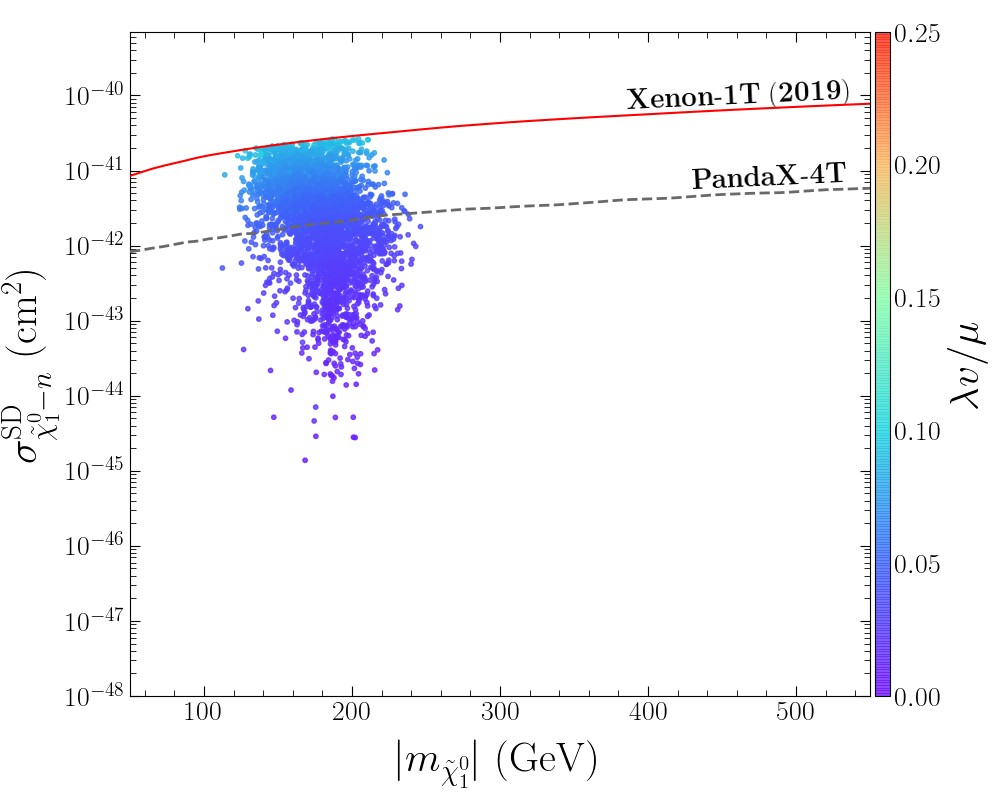}
        }
       \vspace{-0.6cm}
        \caption{Same plot as in Fig. \ref{fig2}, but for the results of the Type-III samples. \label{fig4} }
\end{figure*}

In the following, we investigate the characteristics of the singlino-dominated DM based on numerical results. In Figure~\ref{fig1}, Type-I, -II, and -III samples are projected on the $\sin2\beta-m_{\tilde{\chi}_1^0}/\mu_{eff}$ and $\lambda v / \mu_{eff}-Z_h$ planes. In Figures~\ref{fig2}, \ref{fig3}, and \ref{fig4}, Type-I, -II, and -III samples are projected on the $\sigma_{\tilde{\chi}_1^0-p}^{SI}-m_{\tilde{\chi}_1^0}$ and $\sigma_{\tilde{\chi}_1^0-n}^{SD}-m_{\tilde{\chi}_1^0}$ planes, respectively, with different colors indicating the value of $\lambda v / \mu_{eff}$. In Table~\ref{table2}, a selection of benchmark points are shown to further clarify features in each of the three scenarios.

The following points about Type-I samples can be determined from Figures~\ref{fig1} and \ref{fig2} and the point $P_1$ in Table~\ref{table2}.
\begin{itemize}
\item Type-I samples are characterized by a relatively large $\lambda v/\mu_{eff}$ ranging from $0.14$ to $0.21$ (see right panel of Figure~\ref{fig1}). This will increase the higgsino composition in $\tilde{\chi}_1^0$ through Eq. (\ref{eq:Higgsino/singlino}) and the $\tilde{\chi}_1^0 \tilde{\chi}_1^0 Z$ coupling through Eq. (\ref{eq1:zchi10chi10_S}). As indicated in the right panel of Figure~\ref{fig2}, the SD scattering rates are thus larger than $7 \times 10^{-42}\ {\rm cm^2}$, which are exceedingly close to the near-future PandaX-4T exclusion limit.

\item As shown in the left panel of Figure~\ref{fig1}, $m_{\tilde{\chi}^0_1}/\mu_{eff}$ and $\sin2\beta$ are correlated by $m_{\tilde{\chi}^0_1}/\mu_{eff} \simeq \sin2\beta$.
In this case, $C_{\tilde {\chi}^0_1 \tilde {\chi}^0_1 H_{\rm SM} }$ in Eq. (\ref{eq1:hichi01chi01_S}) is suppressed by $m_{\tilde{\chi}^0_1}/\mu_{eff}$ and $\sin2\beta$ cancellation, favored by the stringent bound of the XENON-1T experiment on the SI cross section. We further explore its implication by focusing on the point $P_1$ in Table~\ref{table2}, which predicts the following four terms in Eq. (\ref{eq1:hichi01chi01_S}):
\begin{eqnarray}
C_{\tilde {\chi}^0_1 \tilde {\chi}^0_1 h} &\simeq& 0.2702 \times (0.6178 - 0.7647 ) + 0.0-0.0029+0.0465 \sim 0.0027,
\label{res_C_xsxshsm1}
\\
C_{\tilde {\chi}^0_1 \tilde {\chi}^0_1 h_{s}} &\simeq& -0.006-0.0+0.0183-0.312 \sim -0.2997.
\label{res_C_xsxshs1}
\end{eqnarray}
The two contributions in Eq. (\ref{eq0:au-ad_hsm-hs}) are as follows:
\begin{eqnarray}
\mathcal{A} & \simeq ~0.0027 -0.0036 \sim -0.0009~.
\end{eqnarray}
These results show that, besides the mentioned cancellation, there is a strong offsetting effect between the first and fourth terms within the $\tilde{\chi}^0_1 \tilde{\chi}^0_1 h$ coupling itself (i.e., cancellation between  $C_{\tilde {\chi}^0_1 \tilde {\chi}^0_1 H_{\rm SM}}$ and $C_{\tilde {\chi}^0_1 \tilde {\chi}^0_1 H_{\rm S}}$ terms in $C_{\tilde {\chi}^0_1 \tilde {\chi}^0_1 h}$) and a strong cancellation between the two contributions to the SI cross section from $h$ and $h_{s}$. These accidental cancellations result in $\sigma^{SI}_{\tilde{\chi}_1^0-p} \sim 10^{-49}\ {\rm cm^2}$. In contrast,  $\sigma^{SI}_{\tilde{\chi}_1^0-p}$ would be around $10^{-42}\ {\rm cm^2}$ without them. This feature explains the SI cross section for Type-I samples possibly being as low as $10^{-50}\ {\rm cm^2}$, as shown in the left panel of Figure~\ref{fig2}.

\item Based on $m_{\tilde{\chi}_1^0}/\mu$, $\lambda v/\mu$, $\sin 2 \beta$, and $Z_h$ in Figure~\ref{fig1}, it is usually predicted that $C_{\tilde {\chi}^0_1 \tilde {\chi}^0_1 G^0} \sim 0.1$ by
Eq. (\ref{eq1:G0chi10chi10_S}) and $\langle \sigma v \rangle_{x_F}^{t\bar{t}} \sim 10^{-26}\ {\rm cm}^3{\rm s^{-1}}$ by Eq. (\ref{eq:gxxGforOmega}). This implies that $\tilde{\chi}_1^0 \tilde{\chi}_1^0 \to t \bar{t}$ played a significant role in determining the abundance. Concerning the point P1 in Table~\ref{table2}, we found $C_{\tilde{\chi}_1^0 \tilde{\chi}_1^0 G^0} \simeq -0.108$ and $\langle \sigma v \rangle_{x_F}^{t\bar{t}} \simeq 1.2 \times 10^{-26}\ {\rm cm}^3{\rm s^{-1}}$, indicating that the annihilation contributed to the total annihilation rate by about $52\%$. We also obtained $\langle \sigma v \rangle_{x_F}^{h_s a_s} \simeq 3.4 \times 10^{-27} \ {\rm cm}^3{\rm s^{-1}}$ by Eq. (\ref{hsaa-approximation}), which means that $\tilde{\chi}_1^0 \tilde{\chi}_1^0 \to h_s a_s$ contributed to the total rate by $15\%$. We add that
our estimation roughly agrees with the results in Table~\ref{table2}, calculated by the \texttt{micrOMEGAs} package.
\end{itemize}

Next we consider Type-II samples. Figures~\ref{fig1} and \ref{fig3} demonstrate following features:
\begin{itemize}
\item $\lambda v/\mu_{eff} \lesssim 0.06$ and $-0.95 \lesssim m_{\tilde{\chi}^0_1}/\mu_{eff} \lesssim -0.90$, which imply that
\begin{eqnarray}
&& |C_{\tilde{\chi}_1^0 \tilde{\chi}_1^0 Z}| \lesssim 2.4 \times (\lambda v/\mu_{eff})^2 < 0.009, \nonumber \\
&& \sigma^{SD}_{\tilde{\chi}_1^0-n}/{\rm cm^2} \lesssim 1.3 \times 10^{-36} \times  (\lambda v/\mu_{eff})^4  < 1.7 \times 10^{-41}
\end{eqnarray}
by Eq. (\ref{eq1:zchi10chi10_S}) and (\ref{eq:sigSD}), respectively. This feature is shown on the right panel of Figure~\ref{fig3}.

\item As indicated in the left panel of Figure~\ref{fig1}, $m_{\tilde{\chi}^0_1}/\mu_{eff}$ and $\sin 2 \beta$ are of opposite sign, and thus,
their contributions to $C_{\tilde {\chi}^0_1 \tilde {\chi}^0_1 H_{\rm SM} }$ in Eq. (\ref{eq1:hichi01chi01_S}) do not cancel each other. This implies that both
$C_{\tilde {\chi}^0_1 \tilde {\chi}^0_1 h }$ and $\mathcal{A}$ in Eq. (\ref{eq0:au-ad_hsm-hs})
are mainly contributed by
$C_{\tilde {\chi}^0_1 \tilde {\chi}^0_1 H_{\rm SM} }$.
Consequently, $\sigma^{SI}_{\tilde{\chi}_1^0-p}$ in Eq.(\ref{eq:SIDD_p2}) is approximated by
\begin{eqnarray}
\sigma_{\tilde {\chi}^0_1-{p}}^{\rm SI}  &\simeq& 5 \times 10^{-45} {\rm cm^2}\times \left(\frac{C_{\tilde{\chi}^0_1 \tilde{\chi}^0_1 H_{\rm SM}}}{0.1}\right)^2,
\end{eqnarray}
where $C_{ \tilde {\chi}^0_1 \tilde {\chi}^0_1 H_{\rm SM}} \simeq (4--10) \times \sqrt{2} \lambda (\lambda v/\mu_{eff})$\footnote{It is noticeable that the enhancement coefficient $4--10$ comes from the factor $1/\{1-(m_{\tilde{\chi}_1}/\mu_{eff})^2\}$ in Eq. (\ref{eq1:hichi01chi01_S}). This is a common characteristic for the Type-II and -III samples.}. This approximation reflects the relation $\sigma_{\tilde {\chi}^0_1-{p}}^{\rm SI} \propto \lambda^4$. For a small $\lambda$, the cross section may be as low as $10^{-48}\ {\rm cm^2}$, as shown on the left panel of Figure~\ref{fig3}.

We add that the SI cross sections in Figure~\ref{fig3} are larger than $3 \times 10^{-49}\ {\rm cm^2}$ because smaller cross sections require smaller values of $\lambda$, which are not readily available since Bayesian evidence is suppressed significantly. We also add that the characteristics of $\sigma_{\tilde {\chi}^0_1-{p}}^{\rm SI}$ for the Type-I and -II samples are different because $\sigma_{\tilde {\chi}^0_1-{p}}^{\rm SI}$ has no significant cancellation effect in the latter case.
Furthermore, we verified the approximation for $\sigma_{\tilde {\chi}^0_1-{p}}^{\rm SI}$ by considering the point $P_2$ in Table~\ref{table2}. The four terms in Eq. (\ref{eq1:hichi01chi01_S}) and the two contributions in Eq. (\ref{eq0:au-ad_hsm-hs}) were as follows:
\begin{eqnarray}
C_{\tilde {\chi}^0_1 \tilde {\chi}^0_1 h}  & \simeq & -0.01 + 0.0 + 0.0 + 0.002 \sim -0.008, \nonumber
\\
\mathcal{A}
&\simeq& ~-0.008 - 0.0008\sim -0.0088.
\end{eqnarray}

\item Since $2 |\kappa|/\lambda \simeq 1$, the singlino-dominated $\tilde{\chi}_1^0$ co-annihilated with the higgsino-dominated neutralinos and charginos to provide the measured abundance. In addition, because $\lambda$ and $\kappa$ must be small to suppress $C_{\tilde {\chi}^0_1 \tilde {\chi}^0_1 G^0}$ and  $C_{\tilde {\chi}^0_1 \tilde {\chi}^0_1 a_s}$, the channels $\tilde{\chi}_1^0 \tilde{\chi}_1^0 \to t \bar{t}, h_s a_s$ never have a crucial effect on the abundance, even if they are kinematically accessible.
\end{itemize}

Finally, we investigate the Type-III samples. Figure~\ref{fig1} indicates that some samples correspond to the same parameter space as the Type-II samples, and thus, they predict
similar DM physics. This conclusion was verified by studying some points in the two scenarios, e.g.,
points P2 and P4 in Table~\ref{table2}. In the following, we only concentrate on the samples with same sign $m_{\tilde{\chi}_1^0}/\mu$ and $\sin 2 \beta$. The following results were obtained from Figures~\ref{fig1} and \ref{fig4}.
\begin{itemize}
\item Compared with the Type-II samples, $\lambda v/\mu_{eff}$ may have a greater value of up to 0.12. The reason is there are accidental cancellations contributing to $\sigma_{\tilde {\chi}^0_1-{p}}^{\rm SI}$, similar to the Type-I samples. This will relax the XENON-1T constraint. We show this characteristic by considering the point $P_3$ in Table~\ref{table2} and finding
\begin{eqnarray}
C_{\tilde {\chi}^0_1 \tilde {\chi}^0_1 h}
 &\simeq& 0.047 \times (0.88 - 0.16) + 0.0-0.0 +0.0085 \sim 0.0437, \nonumber \\
\mathcal{A} & \simeq & 0.042 - 0.0285\sim 0.0135~. \nonumber
\end{eqnarray}
The cancellation also explains why $\sigma_{\tilde {\chi}^0_1-{p}}^{\rm SI}$ was as low as $10^{-50} {\rm cm^2}$, as shown in the left panel of Figure~\ref{fig4}.

\item Due to a relatively large $\lambda v/\mu_{eff}$ and  $| m_{\tilde{\chi}^0_1}/\mu_{eff} |\rightarrow 1$ in Eq. (\ref{eq1:zchi10chi10_S}), $\sigma_{\tilde {\chi}^0_1-{n}}^{\rm SD}$ may be on the border of $10^{-41}\ {\rm cm^2}$ (see right panel of Figure~\ref{fig4}). Furthermore, similar to the Type-II samples, $\tilde{\chi}_1^0$ co-annihilated with the higgsino-dominated neutralinos and charginos to achieve the measured abundance.

We add that the point $P_3$ predicts a significantly larger $\lambda$ than points $P_2$ and $P_4$. Consequently, this will be readily tested in the near-future PandaX-4T experiments for both SI and SD scattering.
\end{itemize}

\section{Conclusion}\label{sec:conclusion}

This work provided updates to previous studies of singlino-dominated DM mainly in four aspects:
\begin{itemize}
\item[(1)] We no longer require the fine-tuning measurement ${\Delta}m_Z$ as a model selection criterion, since it may only reflect personal prejudice.
\item[(2)] We adopted an advanced MultiNest algorithm to perform a sophisticated scan over the $Z_3$-NMSSM parameter space to ensure that the obtained conclusions were as complete as possible.
\item[(3)] We utilized the latest experimental results to restrict singlino-dominated DM scenarios, including the XENON-1T search for both SI and SD DM-nucleon scattering since 2018, ATLAS analyses of sparticle signals with 139-${\rm fb^{-1}}$ data, and measurements of the Higgs couplings with 80-${\rm fb^{-1}}$ data.
\item[(4)] We provided simplified analytical formulas for both the DM annihilation cross sections and the SI and SD cross sections of DM-nucleon scattering, and we numerically scrutinized each contribution in these formulas.
\end{itemize}
As a result, a new singlino-dominated DM scenario (Type-I samples) was found. More model information, such as its Bayesian evidence, was obtained, and the current and future statuses of the scenarios were presented. More importantly, this study provided clear insight into singlino-dominated DM scenarios and explained why they had been tightly limited in the $Z_3$-NMSSM.

Specifically, this study indicated that the surviving samples can be categorized into three types:
 \begin{itemize}
   \item For Type I samples, $0.4\lesssim\lambda\lesssim0.7$,  $0.13 \lesssim \kappa \lesssim 0.23$, $1.5 \lesssim \tan\beta \lesssim 6 $, $450~{\rm GeV} \lesssim \mu_{eff} \lesssim 720~{\rm GeV}$, and the annihilation $\tilde{\chi}_1^0 \tilde{\chi}_1^0 \to t \bar{t}$ is primarily responsible for the DM abundance.
   \item For Type II and III samples, $0 < \lambda \lesssim 0.15$, $\lambda \simeq 2 |\kappa|$, and the dominant annihilation involves a co-annihilation with higgsinos.
 \end{itemize}
The Bayesian evidence ($Z$) for the three sample types showed that the experiments slightly preferred Type-I samples to Type-II and -III. However, Type-I samples will be examined in the
near-future PandaX-4T experiment. They will become highly disfavored if the experiment shows no signs of DM. It should be emphasized that DM-annihilation by a singlet scalar or pseudo-scalar funnel, a Z-boson funnel, and the SM-like Higgs funnel presented by Abdallah et al.~\cite{Abdallah:2019znp} were not observed in this study, due to the small Bayesian evidence. In addition, both analytical formulas and numerical results were used to summarize the theory's four primary cases in significantly suppressing the SI scattering cross section for DM-nucleons. This included 1) a small $\lambda v/\mu_{eff}$, 2) cancellation between $m_{|tilde{\chi}_1^0}/\mu_{eff}$ and $\sin 2 \beta$, 3) cancellation between two contributions from the SM doublet component ($H_{\rm SM}$) and singlet component ($H_s$) within the $\tilde{\chi}^0_1 \tilde{\chi}^0_1 h$ coupling strength itself, and 4) cancellation between two contributions from $h$ and $ h_{s}$.

In summary, the interaction of the singlino-dominated DM with nucleons in the $Z_3$-NMSSM has been tightly restricted in current DM-DD experiments, while the measured abundance favors its involvement in weak interactions. Given the theory's natural preference for electroweak symmetry breaking, it has become increasingly difficult to represent these two seemingly paradoxical features using neutralino DM, due to the limited theoretical structure. Thus, the benefits of singlino-dominated DM are waning unless one extends the $Z_3$-invariant theory. Recent studies on the MSSM and NMSSM imply that DM candidates should be gauge singlet fields or that singlet components should at least be naturally dominant over other components~\cite{Cao:2019qng}. This requirement motivates us to extend the $Z_3$-NMSSM to a general NMSSM to increase the Bayesian evidence of the scenarios significantly~\cite{cao:2021Gnmssm}. It also motivates us to augment the $Z_3$-NMSSM with a seesaw mechanism to generate neutrino masses and select the lightest sneutrino as a DM candidate~\cite{Cao:2017cjf,Cao:2018iyk,Cao:2019qng,Cao:2019aam}\footnote{The properties of the sneutrino DM in the $Z_3$-NMSSM were first studied one decade ago~\cite{Cerdeno:2008ep,Cerdeno:2009dv,Cerdeno:2011qv,Cerdeno:2013oya}. However, these studies considered the cases where the cross section of the sneutrino-nucleon scattering was much larger than current DD experimental bounds, so they obtained different conclusions from our work.}.

\section*{Acknowledgments}
This work was supported by the National Natural Science Foundation of China (NNSFC) under Grant No. 11575053 and No. 12075076.

\section*{Appendix \label{sec:appendix}}

Recently, the ATLAS collaboration aimed to limit the compressed mass spectra case. They analyzed 139 ${\rm fb^{-1}}$ of $\sqrt{s} = 13\ {\rm TeV}$ proton--proton collision data collected at the LHC, focusing on
the events with missing transverse momentum and two same-flavor, oppositely charged, low-transverse-momentum leptons, further categorizing them by the presence of hadronic activity from initial-state radiation~\cite{Aad:2019qnd}. We repeated this analysis using the simulation tools MadGraph5\_aMC@NLO-2.6.6~\cite{mad-1,mad-2} to generate the parton level events, Pythia-8.2~\cite{pythia} for parton fragmentation and hadronization, Delphes-3.4.2~\cite{delphes} for fast simulation of the performance of the ATLAS detector, and CheckMATE-2.0.26~\cite{cmate-1,cmate-2,cmate-3} to implement the analysis cut selections.

Below, we validate our code for all signal regions (SRs)~\cite{Aad:2019qnd}. We considered $\tilde{l}^{+} \tilde{l}^{-}$ production in the MSSM and set the masses of all charginos and neutralinos other than the bino-like $\tilde{\chi}_1^0$ to be 2.5 TeV. Thus, the sleptons will decay by $\tilde{l}^{\pm}\rightarrow \l^{\pm} \tilde{\chi}_1^0 $. We consider the benchmark point $m_{\tilde{l}} = 150~{\rm GeV}$ and $m_{\tilde{\chi}_1^0}=140~{\rm GeV}$. As a result, the cross section at the next-leading order is $126.62~{\rm fb}$ for $\tilde{l}_{L} \tilde{l}_{L}$ production and $47.62~{\rm fb}$ for $\tilde{l}_{R} \tilde{l}_{R}$ production.
The involved cards were set as follows:
\begin{lstlisting}[backgroundcolor=\color{back},frame=trBL]
import model MSSM_SLHA2 --modelname
generate p p > sl1+ sl1-, (sl1- > e- n1), (sl1+ > e+ n1)
add process p p > sl1+ sl1- j, (sl1- > e- n1), (sl1+ > e+ n1)
generate p p > sl2+ sl2-, (sl2- > mu- n1), (sl2+ > mu+ n1)
add process p p > sl2+ sl2- j, (sl1- > mu- n1), (sl2+ > mu+ n1).
\end{lstlisting}
For the proc\_card.dat:
\begin{lstlisting}[backgroundcolor=\color{back},frame=trBL]
100000 = nevents ! Number of unweighted events requested.
0 = ickkw            ! 0 no matching, 1 MLM
37.5  =  ktdurham,
\end{lstlisting}
For the run\_card.dat:
\begin{lstlisting}[backgroundcolor=\color{back},frame=trBL]
Block mass
1000011 1.500000e+02 # Msl1
1000013 1.500000e+02 # Msl2
1000022 1.400000e+02 # Mneu1
1000023 2.500000e+03 # Mneu2
1000024 2.500000e+03 # Mch1
1000025 -2.50000e+03 # Mneu3
1000035 2.500000e+03 # Mneu4
1000037 2.500000e+03 # Mch2
Block selmix
1   1 1.000000e+00 # RRl1x1
2   2 1.000000e+00 # RRl2x2
4   4 1.000000e+00 # RRl4x4
5   5 1.000000e+00 # RRl5x5
\end{lstlisting}
For the param\_card.dat and the pythia8\_card.dat:
\begin{lstlisting}[backgroundcolor=\color{back},frame=trBL]
Merging:Process = pp>{sl1-,1000011}{sl1+,-1000015}{sl2-,1000013}
{sl2+,-1000013}
Merging:mayRemoveDecayProducts=on
\end{lstlisting}
In our simulation, we generated 100,000 events for the production process. The results, shown in Table~\ref{R,L}, indicate that we can reproduce the ATLAS analysis at the $20\%$ level for most cases.
\begin{table*}[t]
	\centering
	\caption{Cut flow for the analysis in~\cite{Aad:2019qnd}. We considered the point $m(\tilde{\ell}, \tilde{\chi}_1^0) = (150, 140)~{\rm GeV}$ in the calculations.}
	\vspace{0.2cm}
	\label{R,L}
    \scalebox{0.75}{
		\begin{tabular}{l|rrrr} 
			\hline
			Process & \multicolumn{4}{c}{Production of $\tilde{\ell}\tilde{\ell}$} \\ \hline
			Point & \multicolumn{4}{c}{$m_{\tilde{\ell}}$ = 150 GeV; $m_{\tilde{\chi}_1^{0}}$ = 140 GeV} \\ \hline
			Generated Events & \multicolumn{4}{c}{100000} \\ \hline
			\multirow{2}{*}{Selection} & \multicolumn{2}{c}{ATLAS} & \multicolumn{2}{c}{CheckMATE}  \\
			& Events & Efficiency & Events & Efficiency \\ \hline
			Total events & 24069 & $\cdots$ & 24069 & $\cdots$  \\
		    $E_{T}^{\rm miss}$ trigger & 2355.37 & $\cdots$ & 2355.37 & $\cdots$  \\
			Two leptons & 1014.55 & 43.07$\%$ & 1079.07 & 45.81$\%$   \\
			veto 3 GeV $< m_{\ell\ell} < 3.2~{\rm GeV}$ & 1013.21 & 99.87$\%$ & 1077.69 & 99.87$\%$    \\
			lepton author 16 veto & 1009.48 & 99.63$\%$ & 1077.69 & 100.00$\%$  \\
			min$(\Delta\phi(\rm any jet, p_{\rm T}^{\rm miss})) >$ 0.4 & 970.36 & 96.12$\%$ & 1049.11 & 97.35$\%$  \\
			$\Delta\phi(\rm j_1, p_{\rm T}^{\rm miss}) >$ 2.0 & 961.15 & 99.05$\%$ & 1027.05 & 97.90$\%$  \\
			lepton truth matching & 958.99 & 99.78$\%$ & 1027.05 & 100.00$\%$  \\
			1 $< m_{\ell\ell} <$ 60 GeV & 827.86 & 86.33$\%$ & 883.55 & 86.03$\%$  \\
			$\Delta R_{\rm ee} >$ 0.3, $\Delta R_{\rm \mu\mu} >$ 0.05, $\Delta R_{\rm e\rm \mu} >$ 0.2 & 826.19 & 99.80$\%$ & 883.48 & 99.99$\%$  \\
			$p_{\rm T}^{j_1} >$ 5 GeV & 823.70 & 99.70$\%$ & 880.95 & 99.71$\%$  \\
			$n_{\rm jet} \geq$ 1 & 810.59 & 98.41$\%$ & 880.95 & 100.00$\%$  \\
			$p_{\rm T}^{\rm j_1} >$ 100 GeV & 705.86 & 87.08$\%$ & 702.58 & 79.75$\%$  \\
			$n_{b-{\rm jet}} = 0$ & 611.05 & 86.57$\%$ & 643.78 & 91.63$\%$  \\
			$m_{\rm \tau\tau} < 0$ or $>$ 160 GeV & 533.29 & 87.27$\%$ & 569.78 & 88.51$\%$ \\
			ee or $\rm \mu\mu$ & 532.33 & 99.82$\%$ & 569.01 & 99.86$\%$  \\ \hline
			\bf SR-highMass & & & & \\
			$E_{\rm T}^{\rm miss} > 200~{\rm GeV}$ & 229.81 & 43.17$\%$ & 265.83 & 46.72$\%$  \\
			max(0.85, 0.98 - 0.02 $\times m_{\rm T_2}^{\rm 100}$) $< R_{\rm ISR} <$ 1.0 & 160.30 & 69.75$\%$ & 165.78 & 62.36$\%$   \\
			$p_{\rm T}^{\ell_2} >$ min(20.0, 2.5 + 2.5$\times (m_{\rm T_2}^{\rm 100} - \rm 100))$ & 70.71 & 44.11$\%$ & 72.51 & 43.74$\%$   \\
			$m_{\rm T2}^{\rm 100} <$ 140 GeV & 70.71 & 100.00$\%$ & 72.51 & 100.00$\%$  \\
			$m_{\rm T2}^{\rm 100} <$ 130 GeV & 70.71 & 100.00$\%$ & 72.51 & 100.00$\%$    \\
			$m_{\rm T2}^{\rm 100} <$ 120 GeV & 70.71 & 100.00$\%$ & 72.31 & 99.73$\%$    \\
			$m_{\rm T2}^{\rm 100} <$ 110 GeV & 70.71 & 100.00$\%$ & 72.23 & 99.90$\%$    \\
			$m_{\rm T2}^{\rm 100} <$ 105 GeV & 53.72 & 75.97$\%$ & 57.10 & 79.05$\%$   \\
			$m_{\rm T2}^{\rm 100} <$ 102 GeV & 20.21 & 37.62$\%$ & 23.77 & 41.63$\%$    \\
			$m_{\rm T2}^{\rm 100} <$ 101 GeV & 9.38 & 46.41$\%$ & 9.90 & 41.62$\%$    \\
			$m_{\rm T2}^{\rm 100} <$ 100.5 GeV & 4.68 & 49.89$\%$ & 4.86 & 49.11$\%$   \\ \hline
			\bf SR-lowMass & & & & \\
			150 $< E_{\rm T}^{\rm miss} <$ 200 GeV & 146.36 & 27.49$\%$ & 167.63 & 29.46$\%$  \\
			0.8 $< R_{\rm ISR} <$ 1.0 & 107.82 & 73.67$\%$ & 93.17 & 55.58$\%$   \\
			$p_{\rm T}^{\ell_2} >$ min(15.0, 7.5 + 0.75$\times (m_{\rm T2}^{\rm 100} - \rm 100))$ & 52.74 & 48.91$\%$ & 42.29 & 45.39$\%$   \\
			$m_{\rm T2}^{\rm 100} <$ 140 GeV & 52.74 & 100.00$\%$ & 42.29 & 100.00$\%$   \\
			$m_{\rm T2}^{\rm 100} <$ 130 GeV & 52.74 & 100.00$\%$ & 42.29 & 100.00$\%$   \\
			$m_{\rm T2}^{\rm 100} <$ 120 GeV & 52.74 & 100.00$\%$ & 42.29 & 100.00$\%$    \\
			$m_{\rm T2}^{\rm 100} <$ 110 GeV & 52.64 & 99.81$\%$ & 41.65 & 98.49$\%$   \\
			$m_{\rm T2}^{\rm 100} <$ 105 GeV & 38.05 & 72.28$\%$ & 29.09 & 69.85$\%$   \\
			$m_{\rm T2}^{\rm 100} <$ 102 GeV & 16.66 & 43.78$\%$ & 11.24 & 38.62$\%$   \\
			$m_{\rm T2}^{\rm 100} <$ 101 GeV & 8.70 & 52.22$\%$ & 5.60 & 49.82$\%$    \\
			$m_{\rm T2}^{\rm 100} <$ 100.5 GeV & 4.39 & 50.46$\%$ & 2.29 & 40.88$\%$   \\ \hline
		\end{tabular}}
\end{table*}

\end{document}